\documentclass[10pt,sigconf]{acmart}
\renewcommand\footnotetextcopyrightpermission[1]{} 
\usepackage{multirow}

\usepackage[flushleft]{threeparttable}

\usepackage{float}
\usepackage{booktabs} 
\usepackage{adjustbox}
\usepackage{algorithmic}
\usepackage[ruled,linesnumbered]{algorithm2e}
\graphicspath{{figure/}{figures/}}

\usepackage{subcaption}
\usepackage{diagbox}
\usepackage{multirow}
\usepackage{color}
\definecolor{circle}{RGB}{47, 135, 151}
\definecolor{comment_changgang}{RGB}{65, 105, 225}
\definecolor{comment_noa}{RGB}{250, 128, 114}

\newcommand\ml{ML\xspace}

\newcommand\pnd{programmable network devices\xspace}
\newcommand\inm{in-network ML\xspace}

\usepackage{enumitem}

\usepackage{xcolor}
\usepackage{colortbl}%
  \newcommand{\grayrow}{\rowcolor[gray]{0.925}}
\usepackage{booktabs}
\usepackage{tikzsymbols}
\usepackage{pifont}
\newcommand{\cmark}{\ding{51}}%
\newcommand{\xmark}{\ding{55}}%
\newcommand{\vmark}{\ding{71}}%

\setcopyright{none}
 
\begin{document}
\title{Automating In-Network Machine Learning}

\author{Changgang Zheng$^{\dagger}$, Mingyuan Zang$^{\S}$, Xinpeng Hong$^{\dagger}$, Riyad Bensoussane$^{\dagger}$, \newline Shay Vargaftik$^\diamond$, Yaniv Ben-Itzhak$^\diamond$, and Noa Zilberman$^{\dagger}$}
\affiliation{
  \institution{$^{\dagger}$University of Oxford, $^{\S}$Technical University of Denmark, $^\diamond$VMware Research}
  \city{$\{$changgang.zheng, xinpeng.hong, noa.zilberman$\}$@eng.ox.ac.uk, minza@dtu.dk, riyad.bensoussane@worc.ox.ac.uk, $\{$shayv, ybenitzhak$\}$@vmware.com}
}

\renewcommand{\shortauthors}{Zheng, et al.}









 \begin{abstract}
 
 Using programmable network devices to aid in-network machine learning has been the focus of significant research. However, most of the research was of a limited scope, providing a proof of concept or describing a closed-source algorithm. To date, no general solution has been provided for mapping machine learning algorithms to programmable network devices.
 In this paper, we present Planter, an open-source, modular framework for mapping trained machine learning models to programmable devices. Planter supports a wide range of machine learning models, multiple targets and can be easily extended. The evaluation of Planter compares different mapping approaches, and demonstrates the feasibility, performance, and resource efficiency for applications such as anomaly detection, financial transactions, and quality of experience. 
 The results show that Planter-based in-network machine learning algorithms can run at line rate, have a negligible effect on latency,  coexist with standard switching functionality, and have no or minor accuracy trade-offs.

\end{abstract}

\settopmatter{printfolios=true}
\maketitle

\section{Introduction}\label{ch1-Introduction}


The rapid growth of data volume and the increasingly heavy demands for data exploitation are creating an ever-increasing processing burden on computing systems~\cite{zhang2017survey}. Concerns about a future shortage of computing resources drove the networking community to consider the underused processing resources within the network~\cite{sanvito2018can,xiong2019switches}. Programmability within the network has been promoted by the emergence of programmable network devices~\cite{feamster2014road,bosshart2013forwarding}. These potential computing resources in the network boost performance, while at the same time increasing power efficiency~\cite{tokusashi2019case}.

\begin{figure}[!t]
	\centering
	\includegraphics[width=0.95\columnwidth]{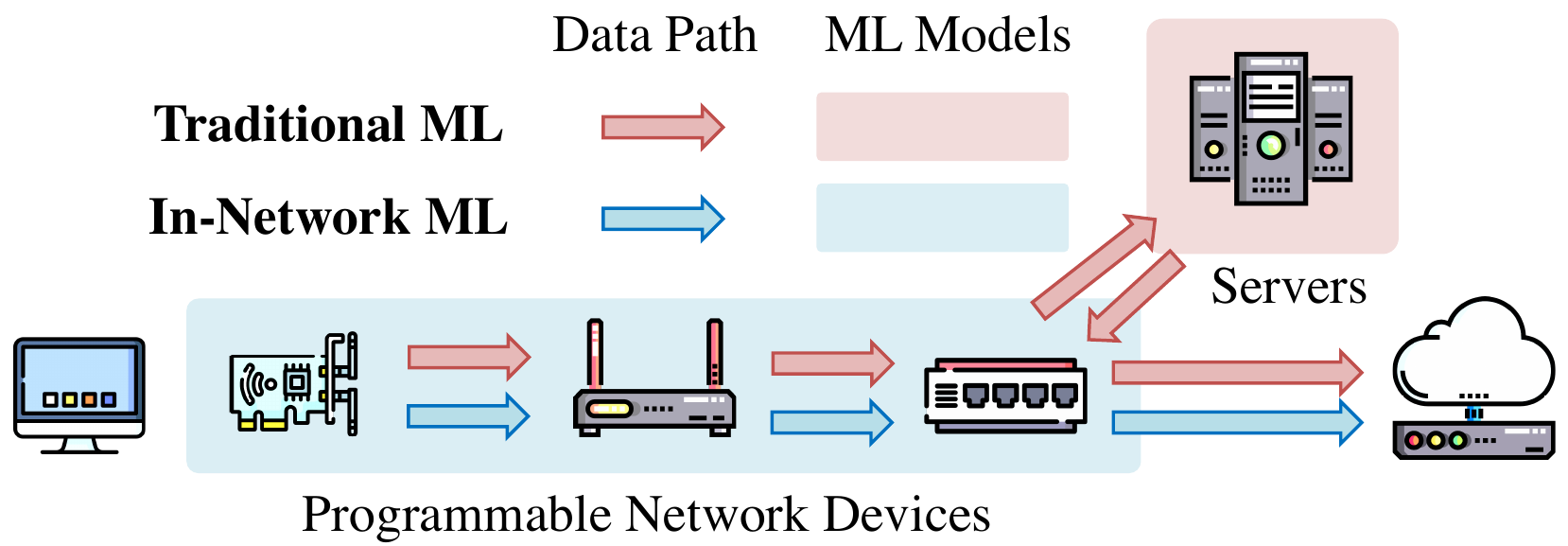}
	\vspace{-1em}
	\caption{Difference in traffic flow between traditional ML in the network domain and in-network ML.}
	\label{fig:in-networkML}
	\vspace{-1em}
\end{figure}

Building upon the programmability of network devices and their high packet processing rate, in-network computing was demonstrated to improve a range of applications, from network services and monitoring~\cite{katta2016hula,macdavid2021p4, ben2020pint, kim2016band, bhamare2019intopt} to caching and consensus~\cite{jin2017netcache, dang2020p4xos}.
These computing tasks require low latency, high throughput and power efficiency, while at the same time flooding the network with data exchanges.
As such, in-network computing was suggested as a means to improve machine learning (ML) performance, both in the acceleration of traditional host-based ML training using in-network aggregation~\cite{sapio2021scaling,lao2021atp}, and ML inference by using in-network ML~\cite{sanvito2018can,xiong2019switches,zheng2022iisy}. 
In-network ML is advantaged by its deployment location, illustrated in Figure \ref{fig:in-networkML}, enabling latency benefits and high throughput. Unlike server-based traditional ML, it does not introduce additional traffic into the network.


Researchers have so far demonstrated the offloading of algorithms such as Support Vector Machines (SVM)\footnote{Appendix~\ref{sec:app_acronym} provides a list of all acronyms used in this paper.}~\cite{xiong2019switches, swamy2020taurus, zheng2022iisy}, Naïve Bayes~(NB)~\cite{xiong2019switches, zheng2022iisy}, K-means~(KM)~\cite{xiong2019switches, friedman2021clustreams, swamy2020taurus, zheng2022iisy}, Decision Trees~(DT)~\cite{zheng2021planter, xiong2019switches,zheng2022iisy}, Random Forest~(RF)~\cite{zheng2021planter, lee2020switchtree, busse2019pforest, xavier2021programmable, zhang2021pheavy, zheng2022iisy}, and (Binary) Neural Networks (NN) \cite{sanvito2018can, zhong2021ioi, siracusano2018deep, swamy2020taurus, siracusano2018network, siracusano2020running, qin2020line}. These works have provided proofs-of-concept and laid the foundation of in-network ML.
However, despite five years of research towards in-network ML, most of the works remained as preliminary implementations, or had prohibitive limitations such as limited model type, size and complexity, compromised accuracy, and lack of comparison to other alternatives. 

Previous in-network ML works have used a wide range of targets, architectures, and mapping procedures. This makes it difficult to reproduce or extend the state-of-the-art, thereby hindering the development and adoption of in-network ML. To that end, a framework that integrates the development and evaluation of commonly-used in-network ML models is needed. 





In this paper, we present Planter, a framework for plug-and-play in-network ML development and deployment. Plant-er supports a range of ML models (e.g., SVM, XGB, BNN, PCA), multiple architectures (PSA~\cite{PSAspecification}, TNA~\cite{TNA}, v1model \cite{peterson2021software}), multiple targets (Tofino~\cite{tofino2021}, P4Pi~\cite{laki2021p4pi} \& BMv2~\cite{bmv2}), and several use cases (e.g., attack detection, financial transactions).
Planter is a modular and scalable solution, enabling support for future in-network classification models and use cases. It also enables a comparison between different ML mapping methods. 

The main contributions of this paper are:
\begin{itemize} 
    \item Introducing Planter, a modular framework for streamlined deployment of different \inm algorithms across multiple architectures and targets. Plant-er's modular design enables easily adding new models, targets, architectures, and features. 

    \item Generalizing, and implementing a wide range of state-of-the-art \inm algorithms, upgrading several previous \inm implementations, and proposing multiple new \inm algorithms mappings.
   
    \item Evaluating and comparing state-of-the-art and newly proposed \inm algorithms in terms of functionality, resources, scalability, and throughput. The comparison results can be used as a selection and referencing guidelines for future in-network ML research.

    
\end{itemize}



\section{Background}\label{ch2-Background}

This section provides the background and the motivation for Planter, as well as setting guidelines for its design.


\paragraph{Programmable Network Devices} The introduction of programmable network devices has enabled users to create customizable data planes. This was further driven by the introduction of the P4 programming language\cite{bosshart2014p4} and the RMT architecture~\cite{bosshart2013forwarding}. Today, programmable data planes are supported on a range of hardware and software targets, using different architectures (PSA~\cite{PSAspecification}, TNA~\cite{TNA}, v1model~\cite{peterson2021software}, and others).

\paragraph{In-network ML} This paper defines in-network ML as \textit{the partial or full offloading of ML algorithms to run within network devices}. In this work, we limit the scope to the forward classification process of ML algorithms being offloaded to the data plane,  while the training part remains on the host (including accelerators) or in the control plane. 
In-network ML algorithms follow an \textit{offline training, online (in-band) inference} pattern. Feature extraction can be done either by parsing within the data plane or customized headers. A mapped ML model is typically implemented within the Match/Action (M/A) pipeline, and the decision can be stored in a header or turned to an action within the network device.
In-network aggregation~\cite{sapio2021scaling, lao2021atp} is outside this scope. 

\subsection{Motivation}
\label{motivating use case}
In network deployments, programmable network devices provide primarily switching and routing-support functions. These functions require a significant portion of the programmable devices' resources, but often don't exhaust them, as demonstrated by Tofino's \textit{switch.p4}. 
In-network ML tasks can utilize remaining resources, co-existing with mandatory and traditional switch functions (see \S\ref{sec:eval-modelfunc}).


Several use cases are commonly tied with ML for networking, and are applicable to in-network ML too:
\paragraph{Traffic Engineering} The use of ML to improve traffic engineering can allow reducing communication overheads between cloud and edge~\cite{Paolucci2019, Paolucci2021}, improve heavy hitter detection~\cite{zhang2021pheavy} and quality of experience (QoE) prediction~\cite{wassermann2020vicrypt}, and support IoT classification at line-rate~\cite{xiong2019switches}.

\paragraph{Anomaly Detection} An important use case of in-network computing and in-network ML is network security~\cite{xavier2021programmable,lee2020switchtree,p4performance}, allowing early detection and fast mitigation of attacks, potentially preventing distributed attacks.

\paragraph{Financial Transactions} ML models are widely used to perform financial tasks, and stock market forecasting is a significant one of them \cite{choudhry2008hybrid, shen2012stock, leung2014machine, henrique2018stock}. In the meanwhile, hardware and software solutions are designed for accelerating the development of financial applications \cite{leber2011high, lockwood2012low}. In a field where every nanosecond counts, the two-fold goal is to increase prediction accuracy, while minimizing latency. 

\subsection{State-of-the-Art In-Network ML}
\label{state of art work}

The in-network ML algorithms realized to date can be divided into three categories: tree-based models (including decision trees and ensemble models), BNN-based models, and other classic ML models.

\begin{table}[htbp]
	\begin{adjustbox}{width=\columnwidth,center}
		\centering
		\begin{threeparttable}
		\begin{tabular}{lcccccccc}
			\toprule
			Work  & ML Models & Targets & PC\tnote{2}     & MS\tnote{2}    & SC\tnote{2}    & PP\tnote{2}     \\
			\hline
			\grayrow SwitchTree  \cite{lee2020switchtree}   &  RF & BMv2 & \xmark &  \xmark & P &  \xmark \\
			   pForest \cite{busse2019pforest}      & RF & Tofino, BMv2 & \xmark & \xmark & \xmark  & \xmark  \\
	
			\grayrow IIsy  \cite{xiong2019switches,zheng2022iisy}   & SVM, KM, DT  & NetFPGA-SUME &  \xmark& \cmark &  \cmark & \xmark \\
			\grayrow     &   NB & BMv2 &  &   &   & \\
		
			Clustream \cite{friedman2021clustreams} & KM & Spectrum-3 &  \xmark& \xmark & \xmark&  \xmark \\
			
			 \grayrow toNIC     \cite{siracusano2018deep}  & NN & NFP4000 & \xmark & \xmark & \xmark &  \xmark \\
			
			     BaNaNa \cite{sanvito2018can}    & NN & RMT-NIC &\xmark  &  \xmark&  \xmark & \xmark \\
			\grayrow N3IC \cite{siracusano2020running}    & NN &  NetFPGA-SUME & \xmark & \xmark &  \xmark & \xmark  \\
			    Qin   \cite{qin2020line}  & NN & Agilio CX, BMv2 &\xmark  & \xmark & P  & \xmark  \\

			\hline
			\grayrow Planter\tnote{1}       & SVM, KM, DT & Tofino, BMv2 &\cmark  & \cmark &   \cmark& \cmark \\
			\grayrow & XGB, RF, NB &  P4Pi-BMv2 &   &  &    &   \\
				\grayrow & NN, PCA, AE &  P4Pi-T4P4S   &  &   &   &   \\
			\grayrow & IF, KNN &     &  &   &   &   \\

			\bottomrule   
		\end{tabular}
		\begin{tablenotes}
    \item[1] Planter builds upon IIsy~\cite{xiong2019switches, zheng2022iisy}, which supports NetFPGA-SUME. However, as SUME is EoL, Planter currently does not support it.
    \item[2] PC - Peer Comparison Provided, MS - Multiple Solutions Provided, SC - Source Code Available, PP - Plug-and-Play Ability Enabled.
  \end{tablenotes}
		\end{threeparttable}
	\end{adjustbox}
	\caption{State-of-the-art in-network ML solutions. }
	\label{tab:state-of-the-art-inm}
	\vspace{-2em}
\end{table}
\begin{figure*}[t]
	\centering
	
	\includegraphics[width=1\linewidth]{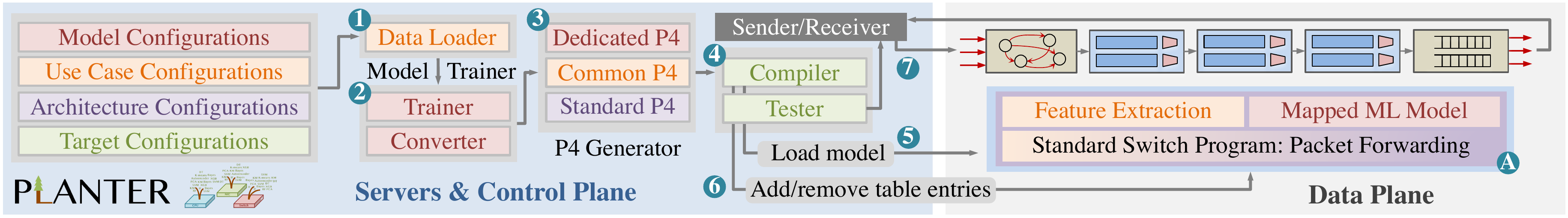}
	\vspace{-2em}
	\caption{The Planter framework components and workflow steps (\textcolor{circle}{\ding{182}} to \textcolor{circle}{\ding{188}}).} 
	\label{fig:framework}
	\vspace{-1em}
\end{figure*}

Table~\ref{tab:state-of-the-art-inm} presents a partial snapshot of recent in-network ML works, with more works discussed in this section. 
Two types of solutions have been presented to tree-based models. The first type, presented in IIsy~\cite{xiong2019switches,zheng2022iisy} is referred to in this paper as an encode-based (EB) solution (see \S\ref{sec:encode}). This solution was also implemented in~\cite{xavier2021programmable}. The second type, used by pForest and SwitchTree~\cite{busse2019pforest, lee2020switchtree}, is direct-mapping (DM) (\S\ref{sec:direct-map}). BNN-based models are mainly based on the XNOR Net\cite{rastegari2016xnor}. With the help of XNOR operations and Hamming Weight (Pop count), Siracusano~\cite{sanvito2018can,siracusano2020running,siracusano2018network,siracusano2018deep,siracusano2022re} and Qin~\cite{qin2020line} realized binarized Multiple Layer Perceptron (MLP) in the data plane. They mainly targeted SmartNICs, which have less rigid limitations than switch ASIC. Other traditional ML models, including SVM, NB, and KM, were introduced by IIsy~\cite{xiong2019switches.zheng2022iisy}, with multiple mappings per model. However, the implementations are FPGA-based and are not resource optimized. Clustreams~\cite{friedman2021clustreams} suggested a different KM mapping of IIsy's.
Taurus \cite{swamy2020taurus} and IOI \cite{zhong2021ioi} use modified ASIC to realize complex operations, and are outside the scope of this paper.

\subsection{The Gap in In-Network ML}
\label{gap}

Although the development of in-network ML solutions is promising, a gap remains between a functional prototype and a plug-and-play solution on commodity devices. As shown in Table \ref{tab:state-of-the-art-inm}, 
 most of the existing in-network ML works support only one type of ML model and one type of a target, typically a SmartNIC or a software switch. A proposed algorithm has only one model mapping solution (in most cases) and is not compared with other in-network ML solutions. Moreover, many solutions have no publicly available source code. Where an artifact was published~\cite{qin2020line,xiong2019switches}, only P4 code is available. Aspects such as weight transformers are missing, and a lot of manual changes are needed when updating a model. 
 
 These aforementioned limitations can be divided into four groups of challenges: high reproduction difficulty, limited model and target options, limited model size, and limited comparison.

\subsection{Planter Design Guidelines}
\label{guideline}
Planter aims to narrow the gap in production in-network ML, and sets the following design guidelines:

\textbf{Ease of use.} 
Mapping ML models to programmable network devices using P4 is not trivial. New targets, architectures, and models may result in significant changes in (i) trained models, (ii) P4 programs, (iii) mapped model parameters to M/A tables or registers, and (iv) table loading process. An easy-to-operate in-network ML end-to-end solution is needed to handle the  process and flexibly adapt to changes (§\ref{ch3-Planter-workflow}).

\textbf{Optimized models.}
Programmable network devices are primarily designed for packet processing and forwarding. They have limited memory and stages, and support a constrained set of mathematical operations and data types. 
Map-ped ML algorithms need to trade off model size and performance to fit on a network device. Thus, the framework should support a wide range of predefined and optimized ML models. 
The resource overhead of the mapped ML algorithms should be minimized as not to affect mandatory network functionality (§\ref{ch6-Machine-Learning}).

\textbf{Modularity.} ML algorithms are emerging rapidly. New architectures and software/hardware targets are reaching the market.
The framework should be able to support new ML algorithms, architectures, and targets. It should be easy to update application scenarios. This calls for a modular design with elements easily added, updated, or replaced, independent of other framework components
(§\ref{ch3-Planter-modular-raw} - §\ref{ch-Target}).

\section{Planter Framework}\label{ch3-Planter}

Planter is a framework for offloading ML models into programmable network devices. The framework uses configuration files to identify the chosen ML model, architecture, target, dataset, and use case. It automatically generates, compiles, loads, and runs the mapped ML models on the target. 

	

\subsection{Workflow and Main Components}
\label{ch3-Planter-workflow}
 
The workflow of Planter, shown in Figure \ref{fig:framework}, has  seven steps. In the first two steps, Planter loads a dataset \textcolor{circle}{\ding{182}} and trains it \textcolor{circle}{\ding{183}}. The  model is mapped to P4 \textcolor{circle}{\ding{184}}, using the selected architecture and target. Generated P4 code is compiled \textcolor{circle}{\ding{185}} and loaded to the data plane target \textcolor{circle}{\ding{186}}. In step \textcolor{circle}{\ding{187}}, table entries and registers are loaded through the control plane. In the final step \textcolor{circle}{\ding{188}}, the auto generated testing module runs a functionality test on the target. 

The generated data plane, shown in Figure \ref{fig:framework} \protect\includegraphics[scale=0.47]{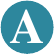}, has three functional parts:  standard switching functionality, feature extraction for ML models, and ML inference. The ML feature extraction and inference can be parallel to the standard functionality (parser operation is merged). 

The workflow is realized using five components: Input Configurations, Data Loader, Model Trainer \& Converter, P4 Generator, and Model Compiler \& Tester. The detailed design of each component is described next. 



\paragraph{Input configurations} Planter is using configuration files to drive its one-click operation.   
The configurations can be loaded from a file, or entered through an interactive CLI. 

\paragraph{Data Loader} The data loader loads datasets for training purposes. It is use-case specific, based on used features and data format. All loaded data are stored in the same format.

\paragraph{Model Trainer \& Converter} ML Training is conducted by the Model Trainer, which drives a standard training framework. Trained models are next mapped to the M/A pipeline in the Model Converter. A software test is generated to test the validity of the mapped model.

\paragraph{P4 Generator} There are three parts to the P4 Generator. The Standard P4 Generator contains architecture-specific P4 code and is the main program that integrates the other P4 codes. This is where the standard network functionality resides. The Common P4 Generator contains the use case specific P4 code, such as bespoke feature extraction. The Dedicated P4 Generator creates the model-related P4 code. 

\paragraph{Model Compiler \& Tester} The Model Compiler \& Tester are deployed in the control plane. The Model Compiler generates bash scripts to compile, load, and run mapped ML models. The Tester generates testing scripts and runs the functionality test on the selected target.\\



\subsection{Modular Framework Design}
\label{ch3-Planter-modular-raw}

Planter is a modular framework. Modules are independent and can be flexibly and easily replaced. The framework supports many ML models,  architecture models, target modules, and use case modules. For navigation simplicity, modules are arranged in folders by type. This can be rearranged by users. In addition to the above, Planter provides a set of common functions, such as exact-to-LPM table conversion, which can be used by other modules. More details are provided in Appendix \ref{ch3-Planter-modular}.

\section{ML Models in Planter}\label{ch6-Machine-Learning}

This section provides a detailed look into the Model Trainer \& Converter component (Figure \ref{fig:framework} step \textcolor{circle}{\ding{183}}). 
Planter supports a range of in-network ML algorithms, e.g., SVM, NB, DT, RF, XGB, IF, KM, KNN, and NN. Among these implemented algorithms, Planter also upgrades some previously proposed implementations (e.g., DT, RF, and NB), and supports new ML algorithms (e.g., XGB, IF, KNN, AE, and PCA). 
The modularity of the framework allows future support in Planter of other types of in-network algorithms, as well as other enhancements.

 \begin{table}[htbp]

	\begin{adjustbox}{width=\columnwidth,center}
		\centering
		\begin{tabular}{lccccccccccc}
			\toprule
			Types  & SVM & DT & RF & XGB & IF & NB & KM  & KNN & PCA & AE & NN\\
			\hline
	
			\grayrow  EB &  &\vmark$_3$  & \vmark$_3$ &  \vmark &\vmark$_2$&  &   \cmark& \vmark &  & & \\

			 LB & \cmark$_3$   &   &  & && \vmark$_2$ &\cmark$_3$ & &\vmark  & \vmark &  \\
			
			\grayrow DM &  &\cmark  &\cmark  & && & && && \cmark \\

			\bottomrule         
		\end{tabular}
	\end{adjustbox}
	\caption{Three types of in-network ML models solutions. Notation: \vmark\ new or upgraded, \cmark reproduced, \cmark$_n$ or \vmark$_n$ $n$ variations exist.}
	\label{tab:planter supported ML algorithm}
	\vspace{-2em}
\end{table}

In Planter, ML algorithms mapping can be classified into three types: encode-based (EB), lookup-based (LB), and direct-mapping (DM). Table \ref{tab:planter supported ML algorithm} shows all the ML models supported under these three approaches. EB solutions encode the feature space for algorithms based on input feature space partitioning. LB solutions are based on lookup in tables of intermediate results. DM approaches map the model directly into the pipeline, using alternative operations or result approximation. 
This section introduces the details of one variation of each model. 
All variations' implementations can be found in \textcolor{black}{\href{https://anonymous.4open.science/r/Planter-Paper-447}{Planter's repository}} \cite{planterrepo}.

 




\subsection{Encode-Based Solutions}\label{sec:encode}

Classification algorithms essentially aim to find borders in a feature space, either the original or a mapped one. The area confined by a set of borders (partitions) is labeled as a class. Algorithms use different methods to define their borders. Some use complex functions, while others use linear functions for approximation. EB solutions mainly use linear borders to slice the feature space with codes representing each part of the area in the space. 

\begin{figure}[htb]
	\centering
	\includegraphics[width=1\columnwidth]{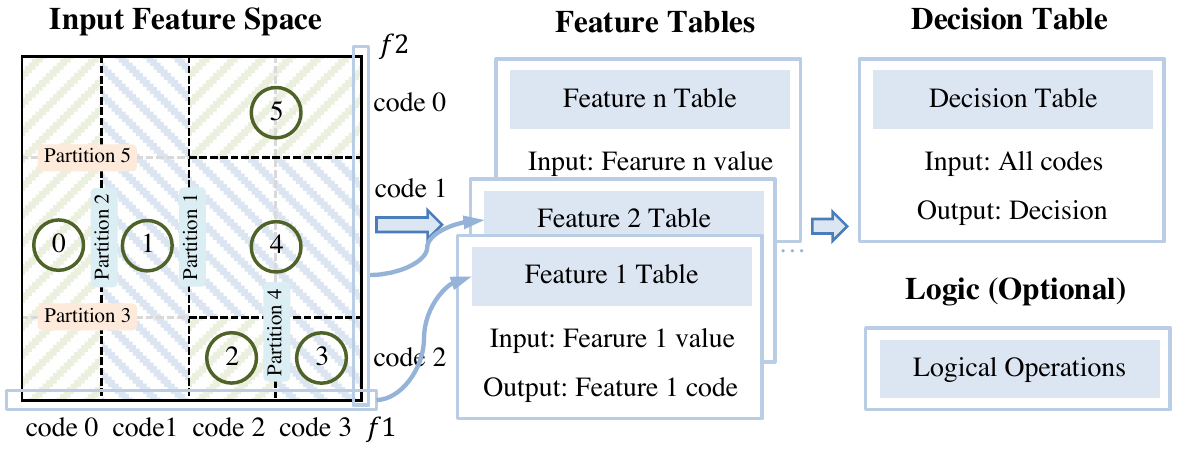}
    \vspace{-2em}
	\caption{Methodology of EB solutions.}
    \label{fig:EB}
	\vspace{-1em}
\end{figure}
\begin{figure*}[htbp]
	\centering
	
	\includegraphics[width=1\linewidth]{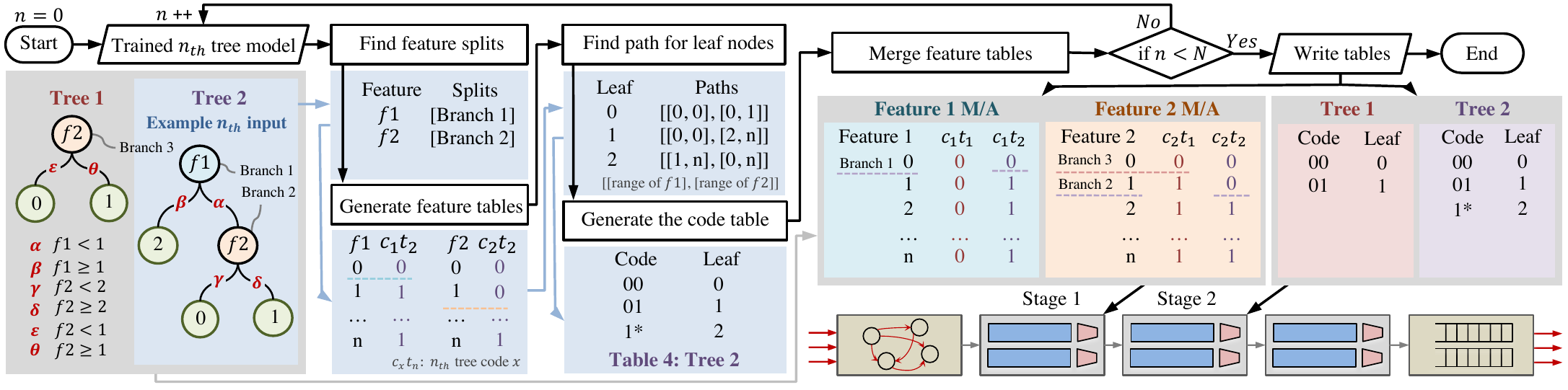}
	\vspace{-2em}
	\caption{Ensemble tree models' workflow using EB solutions.}
	\label{fig:ET workflow}
	\vspace{-1em}
\end{figure*}

To describe the mapping of a general EB model, consider the input features. To slice input features into classes, a typical method uses feature tables and a decision table. As shown in Figure~\ref{fig:EB}, based on a well-trained model, feature space (e.g., two-dimensional space) is sliced into 6 areas (i.e., area \protect\includegraphics[scale=0.47]{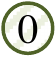} to area \protect\includegraphics[scale=0.47]{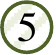}) by 5 partitions (i.e., partition \protect\includegraphics[scale=0.8]{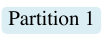} to partition \protect\includegraphics[scale=0.8]{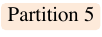}). To map this ML model to M/A pipeline, this input feature space uses two feature tables to record the mapping from feature values to codes, where codes of features represent each area (e.g., area \protect\includegraphics[scale=0.47]{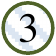} coded as: f1-code 3 \& f2-code 2). The model stores the mapping from codes to labels in a decision (code) table. Similarly, for the case of $n$ dimensional feature space, the model needs $n$ feature tables and 1 decision table. Based on this general method, EB solutions vary depending on how algorithms split the feature space.  

\subsubsection{Decision Tree (DT)}
\begin{figure}[htb]
	\centering
	\includegraphics[width=1\columnwidth]{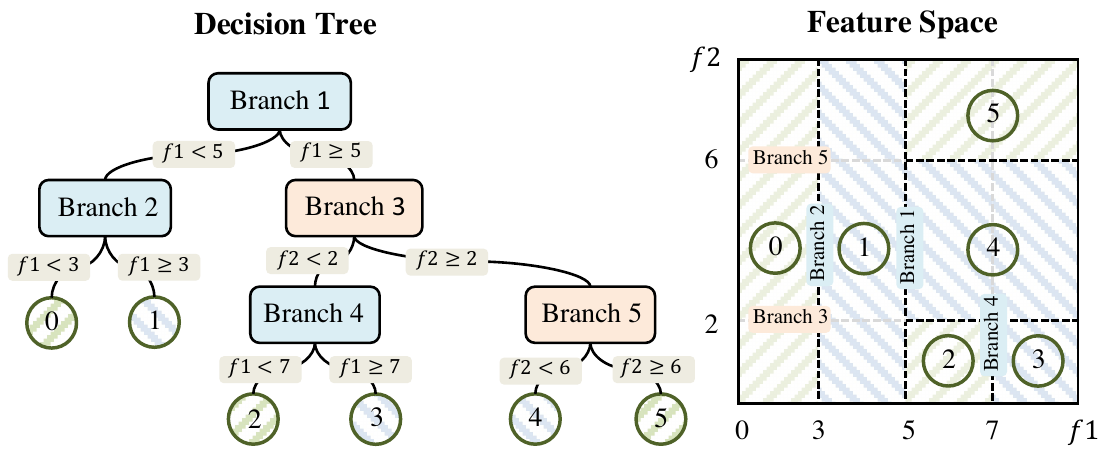}
    \vspace{-2em}
	\caption{Feature space split in decision trees models.}
    \label{fig:DT 1}
	\vspace{-1em}
\end{figure}

DT uses a top-down decision process, and it splits the feature space at each branch (node) until reaching the leaf nodes \cite{wu2008top}. Figure~\ref{fig:DT 1} shows a sample DT model and a two-dimensional input feature space split by its branches.

The similarity between Figures \ref{fig:EB} and \ref{fig:DT 1} shows that a DT model fits the general EB solution. For any $n$ features input space, an encode-based DT (DT$_{EB}$) requires $n$ feature tables, encoding each feature value. The encoded feature space is mapped using a decision table to labels. All feature tables share a pipeline stage (within target limitations) and the entire mapping requires only two logical stages. 


In order to generate features tables and decision tables from a trained model, as shown in Figure~\ref{fig:ET workflow} (illustrating Tree 2 generation process), there are four steps. The process starts with an input of a trained DT. In the step titled ``Find feature splits'', the algorithm collects all the branches related to each feature. Then, feature values are encoded (mapping area to a code word) and saved as feature tables in the step ``Generate feature table''. The encoding is according to the splitting conditions of the branches. The algorithm next looks at each leaf node and determines which part of the feature space belongs to the node and what range of values it includes. Finally, the step ``Generates the tree (also named as code or decision) table''  links the mapping from these leaf nodes to the codes pointing to their pieces.

Different from IIsy's~\cite{xiong2019switches} implementation, the DT$_{EB}$ model uses Ternary match in all feature tables, and uses default actions in a tree table to store the most common label. These improvements can significantly reduce the number of table entries stored in the M/A pipeline, and they are applied in all of Planter's EB ensemble models.

\subsubsection{Random Forest (RF)} RF is an ensemble model built from a set of DT models \cite{ho1995random}. The EB RF (RF$_{EB}$) encodes the trees in parallel and concludes the label by a voting table. 
In the voting table, the RF$_{EB}$ model groups the results of all DT$_{EB}$ constructing the forest as votes in the forward process. Different from the DT model, a key challenge in the RF mapping is looking up multiple trees in parallel. Figure \ref{fig:ET workflow} shows the RF workflow and a toy example of how to map a two-tree RF model to M/A format. For larger models with $n$ feature input and $m$-tree models, the mapped M/A model uses $n$ feature tables and $m$ code tables. Every feature table stores as actions the codes for all trees. In the final stage of the model, voting table are used instead of logic, ao all feature tables and all tree tables are theoretically able to share one logical stage respectively. 

\subsubsection{XGBoost}
XGBoost (XGB) is another type of an ensemble model based on DT. One of the primary differences between XGB and RF is in the value stored at each leaf node. XGB accumulates probabilities from each tree's leaf nodes to make the final decision~\cite{Chen:2016:XST:2939672.2939785}. Due to the operation limitation in some types of programmable devices, it is hard to calculate probablities within the M/A pipeline. To address this issue, the EB XGB (XGB$_{EB}$) encodes all probabilities in each tree. Then, to create the decision (codes-to-label) table, the XGB mapping workflow calculate all combinations of codes and their cumulative probabilities as well as their final label. The probabilities addition and comparison operation thus can be replaced by simple codes-to-label look-ups in the final decision process.

\subsubsection{Isolation Forest}
Isolation Forest (IF) is an unsupervised ensemble model based on RF~\cite{liu2008isolation}. To make the decision, the total number of branches used in the forest decision is compared to an anomaly threshold, as shown Equation \ref{isolation forest}, where $x$ is the input instance, $h(x)$ is the path length, $t$ is the total number of training instance, and $E(h(x))$ is the average $h(x)$ of a collection of trees.
\begin{equation}
 E(h(x)) \leq -(2(ln(t-1)+\gamma)-2(t-1)/t) log_{2}0.5
\label{isolation forest}
\end{equation}
EB IF (IF$_{EB}$) uses a similar method to XGB$_{EB}$ to build the M/A pipeline. The main difference between IF$_{EB}$ and XGB$_{EB}$ is in the M/A table generation process on top of Equation \ref{isolation forest}.
\subsubsection{K-means}


\begin{figure}[htbp]
	\centering
	\includegraphics[width=1\columnwidth]{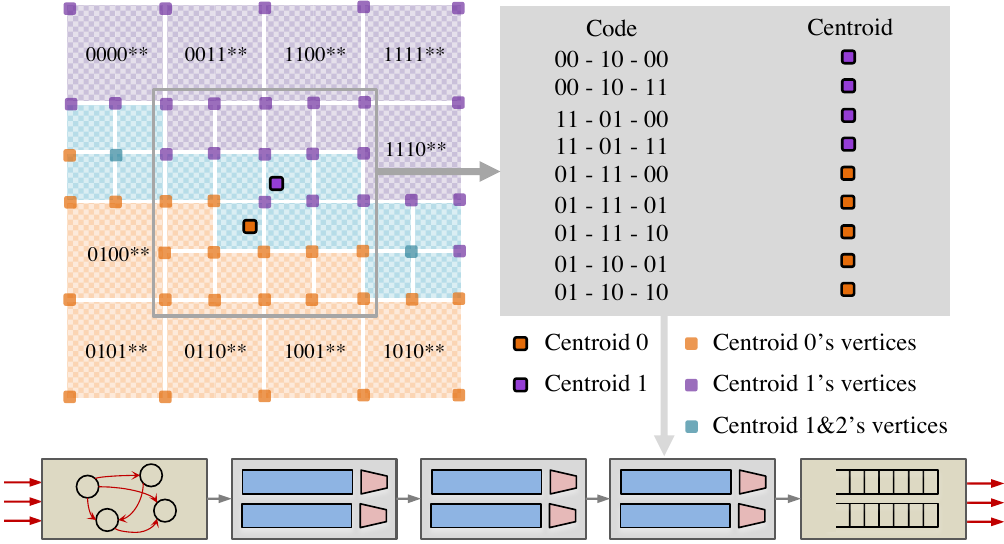}
    \vspace{-1.5em}
	\caption{KM workflow using EB solutions.}
    \label{fig:km1}
	\vspace{-1em}
\end{figure}

The EB KM (KM$_{EB}$) labels the input based on the distance between the data point and each centroid~\cite{fix1989discriminatory}. The input feature space thus can be divided into small pieces. Figure \ref{fig:km1} shows a KM$_{EB}$ toy example in Planter, based on Clustream's~\cite{friedman2021clustreams} solution, which encodes the feature space (2 dimensional) by using the Quadtree. In the higher dimensional feature space, at each depth of the tree, the input $n$ dimensional feature space is divided and labeled into $2^n$ equal parts with the same order. The KM$_{EB}$ requires $d\times n$ bits code to represent each area when the maximum depth is $d$. The feature space is split continuously until the tree reaches the maximum depth or all vertices of the current unit belong to one class. According to its tree-like splitting approach, all these codes are stored in the ternary tables. Compared with the EB tree models, the KM$_{EB}$ requires preprocessing before inference.


\subsubsection{K-nearest Neighbors}
The K-nearest Neighbor (KNN) method splits the feature space in a similar way to KM$_[EB]$. The difference is that KNN uses the distance between the vertices and $k$ nearest neighbors instead of the centroid.

\subsection{Lookup-Based Solutions}
Many ML algorithms involve complex mathematical operations between input features and the final logic. These mathematical operations are commonly too complex to implement in the data plane. Lookup-based solutions use M/A tables to store the intermediate results of these operations and thus are able to realize in-network ML in the data plane.
\begin{figure}[htb]
	\centering
	\includegraphics[width=1\columnwidth]{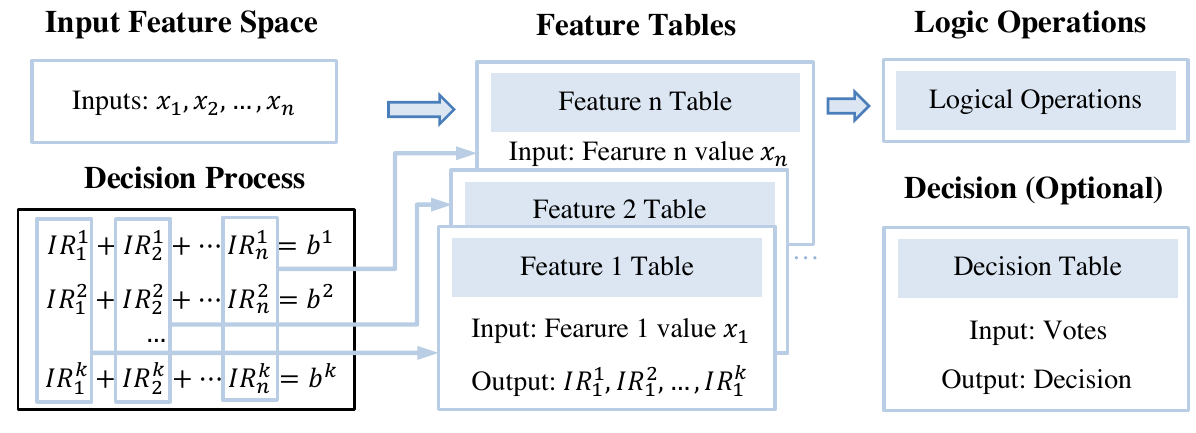}
    \vspace{-2em}
	\caption{Mapping methodology of LB solutions.}
    \label{fig:LB}
	\vspace{-1em}
\end{figure}
Any ML algorithms with a Decision Process can use the LB solutions as shown in Figure \ref{fig:LB}. In LB solutions, feature tables store the mapping between each input feature value and intermediate results. These intermediate values then do the remaining basic mathematical operations, typically addition, as the final stage logic. Multiplication operations between middle results are not performed in the final stage logic, and are only supported through lookup tables.

\subsubsection{Support Vector Machine (SVM)}
SVM maps inputs to hyperplanes, and uses hyperplanes to separate pairs of classes. For a $k$ classification task, SVM model requires $m=k(k-1)/2$ hyperplanes. Each hyperplane is equivalent to a vote. The final vote can be determined by counting the votes from all hyperplanes and using logic or a decision table~\cite{cortes1995support}. 
\begin{equation}
\left\{\begin{array}{c}w_{1}^1 x_{1}+w_{1}^2 x_{2}+\ldots w_{1}^n x_{n}+b_{1}=0 
\vspace{0.5em}\\  w_{2}^1 x_{1}+w_{2}^2 x_{2}+\ldots w_{2}^n x_{n}+b_{2}=0 
\\ \ldots 
\\ w_{m}^1 x_{1}+w_{m}^2 x_{2}+\ldots w_{m}^n x_{n}+b_{m}=0\end{array}\right.
\label{svm-linear}
\end{equation}
For LB SVM (SVM$_{LB}$), IIsy \cite{xiong2019switches} proposed three similar solutions by storing the intermediate result of each hyperplane. The solutions are mainly differentiated by the stored intermediate results. Planter realizes the solution that has the most similar structure with the general solution in Figure~\ref{fig:LB} and with the best scalability, by using $n$ feature tables to store the related middle results from all hyperplanes (e.g., Input: $x_i$, Output $w_{1}^i x_{i}, $ $w_{2}^i x_{i}, $ $\ldots w_{m}^i x_{i}$). Each feature table belongs to a logical stage and uses the addition operation for all hyperplanes (initialized by bias $b_i$). This method has an optimal memory and stage consumption compared to other IIsy approaches.

\subsubsection{Naïve Bayes (NB)}
For every set of inputs, NB calculates the posterior probability of each class~\cite{domingos1997optimality}. As shown in Equation~\ref{bayes forward}, the Bayes model chooses the label that maximizes the probability as the final decision.
\begin{equation}
\hat{y}=\arg \max _{y} P(y) \prod_{i=1}^{n} P\left(x_{i} \mid y\right)
\label{bayes forward}
\end{equation}
IIsy provides one NB solution that belongs to the LB solution, which directly uses all features as the M/A table input and outputs the respective posterior probabilities of all their classes~\cite{xiong2019switches}. This method only works when each input feature has a relatively narrow value domain. Otherwise, the table will be too large for a programmable device. This is because  all intermediate results $P\left(x_{i} \mid y\right)$ are connected through multiplication rather than of addition, and most switch ASICs do not support multiplication.

In Planter, we realize an upgraded LB Bayes model, which uses logarithm operations to convert multiplication into addition, as shown in Equation \ref{bayes forward new}.
\begin{equation}
\hat{y}=\arg \max _{y} [map(log_2P(y)) +\sum_{i=1}^{n} map(log_2P\left(x_{i} \mid y\right))]
\label{bayes forward new}
\end{equation}
The upgraded Bayes model now fits the standard LB solution, shown in Figure \ref{fig:LB}, which uses $n$ feature tables (e.g., Input: $x_i$, Output $map(log_2P(x_{i} \mid y_1)), map(log_2P(x_{i} \mid y_2)), $ $\ldots, map(log_2P(x_{i}\mid y_k))$) for any $k$ classes inference task.

\subsubsection{K-means}

The KM workflow labels inputs according to their distance to the trained $k$ centroid \cite{fix1989discriminatory}, as described in Equation \ref{K-means}. Based on the LB approach, the LB KM (KM$_{LB}$) (IIsy's implementation \cite{xiong2019switches}) workflow can use the standard solution. The workflow uses $n$ feature tables to store the intermediate result in parallel. 
\begin{equation}
D_{i}=\sqrt{\left(x_{1}-c_{1}^{i}\right)^{2}+\left(x_{2}-c_{2}^{i}\right)^{2}+. .\left(x_{n}-c_{n}^{i}\right)^{2}}
\label{K-means}
\end{equation}
The final step of the distance calculation is square root operation. It is hard to do this operation directly in data plane. The square root function is monotonically increasing in a specific domain ($>1$). This final square root step can be ignored when the workflow has no value located outside that specific domain. The KM$_{LB}$ solution thus uses $map(.)$ operation and construct feature table with input $x_i$, output $map(x_{i}-c_{i}^{1}), map(x_{i}-c_{i}^{2}), $ $\ldots,  map(x_{i}-c_{i}^{k})$. The $map(.)$ function maps all input value to a domain $\{1:2^{n_{bits}}/n\}$, where $n_{bits}$ is the width of each action data.

\subsubsection{Autoencoder} Autoencoder workflow is composed of an encoder and a decoder \cite{liou2014autoencoder}. The forward path of a trained encoder is equivalent to a small encode network. The single-layer encoder network has a similar format as the Decision Process in Figure~\ref{fig:LB} and thus can be realized by using the standard LB solution. 
\begin{equation}
X_{new} = XW+B
\label{auto}
\end{equation}
Equation \ref{auto} demonstrates the signal layer encoder. When $X_{new}$ has $k$ dimensions, $W$ is a $n\times k$ weight matrix. The workflow uses $n$ feature tables to store the intermediate result of all output feature dimensions $(x_iw^i_1,$ $ x_iw^i_2,$ $ \ldots,$ $ x_iw^i_k)$ under the corresponding feature $i$. The final logic add all the intermediate results in each output dimension and the bias as the output.

\subsubsection{Principle Component Analysis (PCA)}

PCA finds a new axis with a predefined dimension that can best represent the feature space \cite{doi:10.1080/14786440109462720}. As shown in Equation~\ref{pca}, the forward path of a trained PCA has two main steps: move and map.
\begin{equation}
X_{new} = (X-X_{means})Components
\label{pca}
\end{equation}
In this equation, the input $X$ is the array $[x_1, x_2, ..., x_n]$ with $n$ input features, and the mean of each feature is $X_{means} = [x_{means}^1$, $x_{means}^2$, $...$, $x_{means}^n]$. The $Components$ is a transferring matrix with $n$ rows and $m$ columns. The output $X_{new}$ is the array $[x_{new}^1$, $x_{new}^2$, $...$, $x_{new}^m]$ with $m$ output features. This equation fits the LB solutions. The LB Autoencoder thus uses $n$ feature tables. The intermediate result in feature table $i$ is $IR_i^1 = (x_i-x_{means}^i)w_i^1, $ $IR_i^2 = (x_i-x_{means}^i)w_i^2, $ $\ldots, $ $IR_i^m = (x_i-x_{means}^i)w_i^m$

\subsection{Direct-Mapping Solutions}\label{sec:direct-map}
Some of the ML algorithms have a relatively similar structure to the data plane architecture, which can be deployed into the data plane without significant structural change. However, there are still many operations that should be replaced to meet the data plane architecture before being able to be mapped to the pipeline.

\subsubsection{Decision Tree (DT)}
\begin{figure}[htbp]
	\centering
	\includegraphics[width=1\columnwidth]{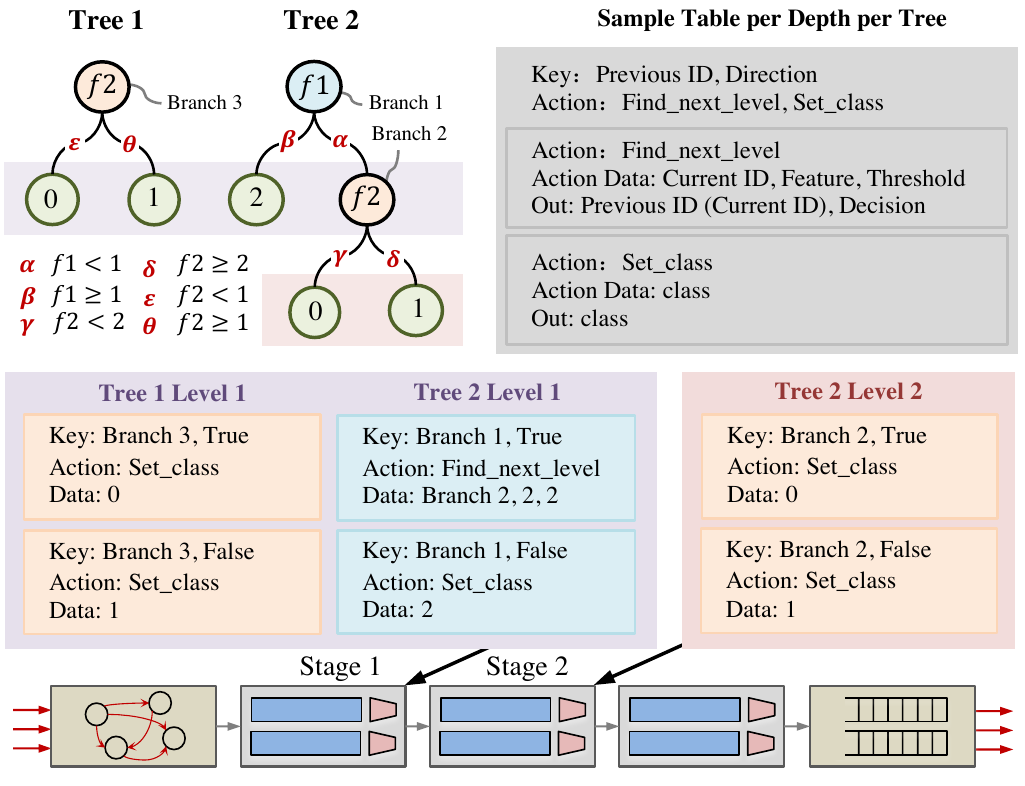}
    \vspace{-2em}
	\caption{Ensemble Tree models workflow using DM methodology.}
    \label{fig:ensemble tree 2}
	\vspace{-1em}
\end{figure}

The workflow of the DT model is similar to the M/A pipeline. In the DM DT (DT$_{DM}$) model, the workflow goes through the nodes from top to bottom, does the value comparison, and finally uses the compared result as keys to find the next layer's branch until reaching the leaf node. As shown in Figure~\ref{fig:ensemble tree 2}, take Tree 2 as an example, the workflow presents how the DT model can be realized in \pnd by using the direct-mapping approach. The DT$_{DM}$ in Planter is based on the work pForest \cite{busse2019pforest} and SwitchTree \cite{lee2020switchtree}. For a $p$ depth DT model, the mapped model can use $p$ depth table. In each table, based on the key from lower depth, the workflow checks the current branch ID, its threshold and the used feature. After the lookup, the process does the comparison based on the threshold and the feature. The comparison result and the current bran ID are used as the keys when the workflow dives into deeper layers.   

The DT$_{DM}$ approach consumes little memory. However, after each lookup, the logic operations are complex. First, the workflow needs to choose which feature value to be used, then compares it with the threshold. These operations are stage-consuming and latency-consuming especially when input feature numbers are large.

\subsubsection{Random Forest (RF)}

RF and any other ensemble tree models can use a similar workflow. As shown in Figure \ref{fig:ensemble tree 2}, the toy example of two trees on the top left corner can be mapped to the data plane via the top right format. The mapped table for the toy example for two trees is represented in the purple and red box respectively. After obtaining all the votes from each tree by using the figure's workflow, the DM RF$_{DM}$ applies a similar decision process as the EB RF$_{EB}$, which uses logic or a decision table. Planter supports the logic version in the end.


\subsubsection{Neural Networks (NN)} Based on the work toNIC \cite{siracusano2018deep} and N3IC \cite{siracusano2020running}, the trained NN can be mapped into the M/A pipeline as BNN via DM approach. The matrix multiplication between each layer's input and weights are replayed by XNOR and PopCount operations \cite{rastegari2016xnor}. Not all programmable network devices can realize this model. 
\begin{figure}[htbp]
	\centering
	\includegraphics[width=1\columnwidth]{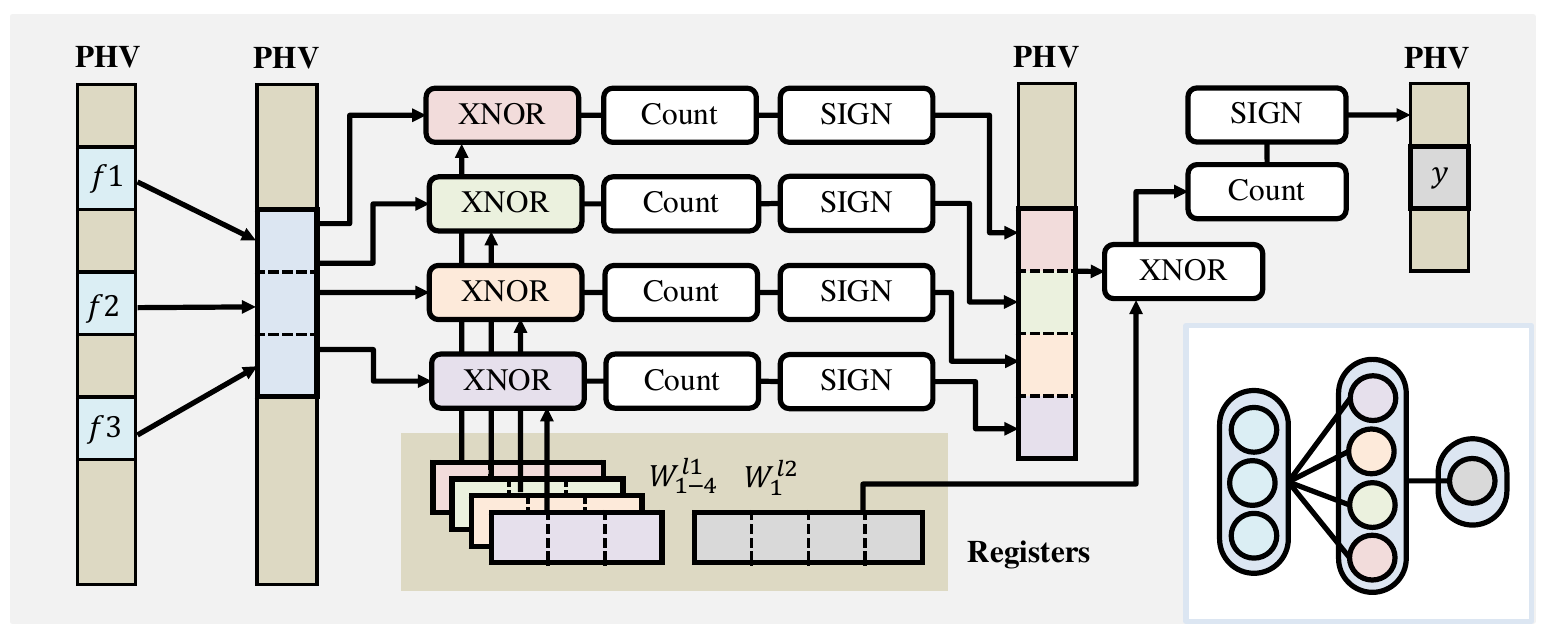}
    \vspace{-2em}
	\caption{BNN workflow using DM methodology.}
    \label{fig:NN}
	\vspace{-1em}
\end{figure}
\begin{equation}
x^{l(i+1)}_j = SIGN(PopCount(XNOR(X^{li}, W^{li}_j)))
\label{NN}
\end{equation}
Figure \ref{fig:NN} shows a NN toy example with one four-node hidden layer based on Planter implementation. After weights binarizing the trained NN or directly using the weight from the XNOR net, the model stores the weight in registers. The DM workflow firstly concatenates the input feature as an input vector. For any input $X^{li}$ for layer $i$, the workflow performs XNOR, PopCount, and SIGN operations between inputs and weight as illustrated in Equation \ref{NN}, where layer $i$ have $k$ nodes with weight $W^{li}_1, W^{li}_2, \ldots, W^{li}_k$. The next layer input is calculated by concatenating the results of all nodes in the current layer ($X^{l(i+1)}= x^{l(i+1)}_1++x^{l(i+1)}_2++\ldots++x^{l(i+1)}_k$). In the final layer, the model can choose to directly output PopCount results without activation.

\section{P4 Architectures and Targets}\label{ch-Architecture}\label{ch-Target}


In the Planter framework, P4 architectures are the shell for the in-network ML models, as shown in Figure \ref{fig:framework} \protect\includegraphics[scale=0.47]{figures/c_A.pdf}. Planter currently supports three architectures: Intel Tofino Native Architecture (TNA)~\cite{TNA}, v1model, and Portable Switch Architecture (PSA)~\cite{PSAspecification}. 
Other architectures, such as Portable NIC Architecture (PNA) for NICs and SimpleSumeSwitch~\cite{ibanez2019p4} for FPGAs are subject to future support. Support for new architectures requires adding architecture-specific Standard P4 module in the P4 generator block (Figure \ref{fig:framework} \textcolor{circle}{\ding{184}}).

P4 targets refer to the software or hardware platform of a data plane, such as a SmartNIC or a Switch ASIC. Planter currently supports Intel Tofino, BMv2, and P4Pi~\cite{laki2021p4pi} using either T4P4S over DPDK or BMv2. Targets can use different architectures. For example, P4Pi supports PSA and simple\_switch\_grpc (based on v1model).  Support of new targets will affect the Model Tester and Compiler (shown in Figure \ref{fig:framework} \textcolor{circle}{\ding{185}}), and requires adding scripts driving the target's compiler and a target testing environment.

\section{Implementation}\label{ch10-Implementation}


The Planter framework is implemented in 17697 lines of code in Python, and is available at \textcolor{black}{\href{https://anonymous.4open.science/r/Planter-Paper-447}{Planter's GitHub repository}}
\cite{planterrepo}. The framework trains a model using a python-based learning framework, such as PyTorch/Scikit-learn. A trained model is mapped into M/A format, and the framework saves table entries and generated weights (for NN) into JSON/txt (target dependent) files. Data plane P4 codes are consequently automatically generated. Planter further generates Bash scripts to interact with the target for model deployment and verification. Control plane support is target dependent, and loading table using P4Runtime is supported.
\begin{table}[htbp]
  \begin{adjustbox}{width=1\linewidth,center}
  \centering
    \begin{tabular}{l c c c c c c c}
    \hline
    Python &Architecture  & Target & Model   & Use Case  &  Data   &   Function     \\
    \hline
    \grayrow Average &  137   & 542   &477 & 91  & 88 & 29   \\
    Maximum &  145& 793  &   665 & 127  & 148 &  130     \\
    \grayrow Minimum &  123   &  291 & 293  & 54  & 16  &  6 \\
    \hline
    \end{tabular}%
  \end{adjustbox}
  \vspace{0.1em}
  \caption{The average lines of codes of each module.}
  \label{tab:lines}%
  \vspace{-2em}
\end{table}%
To illustrate the modularity of Planter, Table \ref{tab:lines} presents the average, minimum and maximum lines of code (LOC) required to support different modules in Planter. ML models require an average of 477 LOC, and no more than 665 LOC. Supporting a new P4 architecture requires less than 150 LOC, and supporting a target requires less than a thousand LOC. This lightweight implementation of architectures, targets, and models is a key enabler and an advantage of Planter. New datasets and use cases may require bespoke parsing or dataset processing, captured under ``Use Case'' and ``Data'' columns. Shared framework functionality, denoted by ``Function'', requires 377 lines in total.

\section{Evaluation}\label{ch11-Evaluation}

The evaluation of Planter focuses on three aspects: inference performance of different in-network ML algorithms, system performance, and model scalability under different scenarios. Our results show: 


\begin{itemize}
   \item[(1)] Most In-network ML algorithms can be implemented on a commercial switch (Tofino), and coexist with L2/L3 switching functions.
    \item[(2)] Planter-generated in-network ML algorithms reach line rate on a commercial switch (Tofino), with negligible change in latency. Over half of the algorithms exceed 80\% of maximum throughput on P4Pi. 
    \item[(3)] Most in-network ML algorithms have the same accuracy as server based models, or have a slight accuracy loss, for the same model size. Large server-side models can lead to small accuracy differences. 
    \item[(4)] For most models, the framework runtime is less than 10 seconds.
    \item[(5)] The scalability of some algorithms is independent of hyperparameters and use case, while in others there is an increase in resource consumption.
\end{itemize}

\subsection{Methodology and Testbed Setup}
\textbf{Testbed setup: } Our testbed uses a Tofino switch (APS-Networks BF6064X), a server (ESC4000A-E10, AMD EPYC 7302P CPUs, 256GB RAM, Ubuntu 20.04LTS), and a Raspberry Pi 4 Model B with 8GB RAM. The Raspberry Pi set as P4Pi running v1model over BMv2 software switch. The Tofino switch uses a snake configuration for throughput tests. More details in Appendix~\ref{sec:app_setup}.



\textbf{Workloads: } Our evaluation explores four use cases: attack detection (using AWID3 \cite{chatzoglou2021empirical}, CICIDS 2017 \cite{sharafaldin2018toward}, KDD99 \cite{kddfeatures}, and UNSW-NB15 \cite{moustafa2015unsw}), finance (NASDAQ TotalView-ITCH \cite{nasdaqnasdaq}, Jane Street Market Prediction \cite{janekaggle}), QoE (Requet \cite{requetdataset}) and flowers classification (Iris \cite{fisher1936useIrisDataset}). The results for attack detection (using CICIDS and UNSW) and finance are presented below, and the rest are described in appendix \ref{apdix: eval-add-dataset}.
The attack detection use case uses 5 features: Source IP, Destination IP, Source Port, Destination Port, and protocol (KDD uses duration, protocol\_type, service, flag, and land). We use three packet-level fields (order side, size, and price) as features in NASDAQ dataset and five packet-level features (stock market data 42, 43, 120, 124 and 126) in Jane Street Market dataset. In this manner, the evaluation explores both stateless feature extraction (attack detection) and stateful features (finance).

\textbf{Parameter settings: } Mapped in-network ML models are explored using four different model sizes: small (S), medium (M), large (L), and huge (H). Parameters' setting per use case are provided in Appendix~\ref{Evaluation Details} Table \ref{tab:parameters_settings}. The model size refers to the converted data plane model size, which is a function of both training and conversion parameters. Small to large in-network ML models are expected to fit on the target data plane. Huge models represent the maximum inference potential of each type of model per dataset.

\textbf{Evaluation metrics: }
The following metrics, explained in Appendix \ref{Evaluation Details}, are used in the evaluation:
\begin{enumerate}
    \item Inference performance: \textit{Accuracy}, \textit{F1 score} and \textit{Pearson correlation coefficient} are used to evaluate the inference performance of ML algorithms.   
    \item System performance: \textit{Throughput} and \textit{latency} are used to evaluate the system performance of mapped models.   
    \item Model scalability: \textit{Memory utilization, table entries}, and \textit{number of stages} are used to evaluate scalability. 
    \item Framework performance:  \textit{model training time} and  \textit{train-ed model conversion time} are used to assess Planter's run time performance.
\end{enumerate}
On Tofino, following NDA, we record the memory utilization and latency relative to $switch.p4$ reference switch program.

\subsection{Framework Execution Time}


\begin{figure}[htbp]
	\centering
	\includegraphics[width=1\columnwidth]{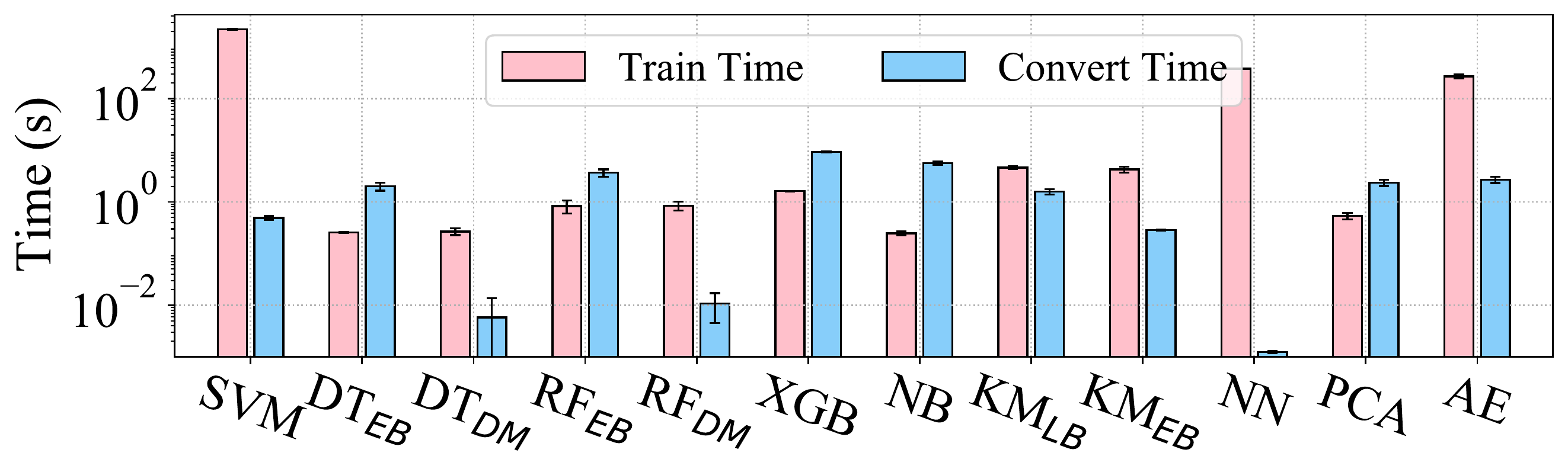}
    \vspace{-2.5em}
	\caption{Train and convert time of ML algorithms.}
    \label{fig:Model_name-Times-S}
	\vspace{-1em}
\end{figure}

We measure the time required to load a dataset, train a model, convert the trained model, test table entries, compile the mapped model to a target, and load the generated tables. Among these, we focus on training and conversion time, the two time-consuming components in Planter's functionality. As shown in Figure~\ref{fig:Model_name-Times-S}, for a small model, most of the models' training time (except SVM, NN, and AE) and all of the models' conversion time are less than $10S$. Due to their intrinsic structure and mapping algorithm, the conversion time of trained XGB and KM$_{EB}$ models is sensitive to the model size more than other models. The conversion time of these two models noticeably increases when using a medium size model (Appendix \ref{Evaluation Details} Figure \ref{fig:Model_name-Times-M}).

\begin{table*}[htbp]
  \begin{adjustbox}{width=1\linewidth,center}
 
\centering
\begin{tabular}{ccccccccccc||cccccccccccc} 
\hline\hline
                          & \multicolumn{10}{c||}{Accuracy}                                                                                                                                                                         & \multicolumn{12}{c}{System Performance}                                                                                          \\ 
\hline
\multicolumn{1}{l}{}      & \multicolumn{4}{c}{CICIDS}                                                                        & \multicolumn{6}{c||}{UNSW}                                                                          & \multicolumn{12}{c}{UNSW}                                                                                                        \\ 
\hline
\multicolumn{1}{l}{}      & \multicolumn{2}{c}{Switch (M)} & \multicolumn{2}{c}{Sklearn (M)}   & \multicolumn{2}{c}{Switch (M)} & \multicolumn{2}{c}{Sklearn (M)} & \multicolumn{2}{c||}{Server (H)} & \multicolumn{3}{c}{ACC (Switch)}       & \multicolumn{3}{c}{Memory (Relative)} & \multicolumn{3}{c}{Latency (Relative)} & \multicolumn{3}{c}{Stage (Tofino)}  \\ 
\hline
\multicolumn{1}{l}{Model}    & ACC   & F1                      & ACC   & F1                     & ACC   & F1                     & ACC   & F1                      & ACC   & F1                      & S      & M     & L                       & S      & M     & L         & S      & M     & L          & S      & M      & L        \\ 
\hline
\grayrow SVM                       & 59.24     & 37.20                         & 95.04    & 94.94                    &  97.31   &    49.32       &  99.23  &       93.51     & 99.23  &       93.51             & 97.31 & 97.31 & 99.23                   & 4.13 & 5.57 & 7.09   & 26.37 & 35.27 & 35.30     & 9 &9 & 9    \\
DT$_{EB}$                      & 99.92     & 99.92             & 99.92     & 99.92            &   99.40 & 94.53             & 99.40 & 94.53                   &  99.40 &     94.31               & 99.34      & 99.40    & 99.41                       & 2.27      & 2.54     & 2.54         & 26.37      & 26.37     & 26.37          & 4      & 4      & 4       \\
\grayrow DT$_{DM}$                       & 99.92     & 99.92  &  99.92     & 99.92  &   99.40 & 94.53                & 99.40 & 94.53      & 99.40  & 94.31                 & 99.34     &99.40     & 99.41                       & 2.51      & 2.96     & 3.41       & 81.16      & 88.36    & 88.36          & 11      & 13\dag      & 15\dag        \\
RF$_{EB}$                      & 99.80    & 99.79             & 99.80    & 99.79                        &    99.37 &      94.41         &99.38  &    94.44               &  99.42 & 94.51                      & 99.25     &  99.37    & 99.39       & 3.17     & 4.79     &    7.11      & 39.04      & 39.40    & 45.89         & 5      & 5      & 8        \\
\grayrow RF$_{DM}$                       &99.80    & 99.79                        & 99.80    & 99.79                         &  99.38  &    94.44              &  99.38  &    94.44               &  99.42 & 94.51         & 99.25      & 99.38     &      99.39   &13.29      &24.15     & NF        & 88.36     & 89.04     & NF         & 41\dag     & 77\dag      & NF        \\
XGB                       &99.98     & 99.98                     & 99.98     & 99.98                               &  99.42 & 94.53             &  99.42 & 94.53                    & 99.43 &    94.59                 & 99.40      & 99.42     &    99.45   & 6.40     & NF     & NF      & 33.22      & NF     & NF        & 7      & NF      & NF       \\
\grayrow IF &   44.89  &  35.35   &    37.90   &  31.08   &    84.86   &   58.90     &   63.83   &  45.07     &  86.33  &  55.05    &    81.74   &    84.86    &   NF  &   7.24   &   9.01    &  NF    &    36.30  &     43.33 &     NF & 5  &  5 & NF   \\
NB                        & 98.99   &   98.95      &   98.99   &   98.96      &    99.25   &        93.68               & 99.25   &        93.68       & 99.25   &        93.68        & 99.25     &   99.25     & 99.25                       & 5.66      & 7.27     & 10.70         & 28.77      & 28.77     & 28.77          & 8     & 8     & 8       \\
\grayrow KM$_{LB}$                       &  58.40 & 56.80     &         58.40    &    56.80         &       71.28    & 41.88   &  71.28    & 41.88      & 71.28    & 41.88        &   71.55  &    71.28 &               71.28           & 5.37      &   6.82    & 9.96         & 21.58      &  21.58   & 21.58          & 7     &   7     & 7       \\
KM$_{EB}$                       & 56.92  &  55.75       & 58.40 & 56.80                 &   72.69  &    42.37     &  71.28    & 41.88        &   71.28    & 41.88          &    77.21   &  72.69    &       71.30                & 0.78      &   6.26   &  NF       & 19.52      & 19.52     & NF         & 2      & 2    & NF        \\

\grayrow KNN &      69.33    &      60.63  &   99.38    &     99.36 &  87.51     &  31.55  &    99.30  & 93.17    &  99.30  & 93.17   &   78.24     &   87.51    &   92.73   &  0.64   &    5.55   &     62.18 & 20.74  &    20.74  &  22.22    & 1  &  1  &  5  \\
NN                        &     92.09     &     92.00    & 99.96 & 99.96                         &    98.33   & 85.68                      & 99.25     & 93.67                       & 99.25     & 93.68                       & 98.33       & 98.33    & 97.50                     & NF     & NF    & NF        & NF     & NF    & NF         & NF    & NF     & NF       \\ 
\cline{1-14}
~                  & P1    & P2                      & P1    & P2                     & P1    & P2                     & P1    & P2                      & P1    & P2                      & P1     & P1    & \multicolumn{1}{c}{P1} & \textcolor{white}{00.00}      & \textcolor{white}{00.00}      & \textcolor{white}{00.00}          & \textcolor{white}{00.00}       & \textcolor{white}{00.00}      & \textcolor{white}{00.00}         & \textcolor{white}{000}       & \textcolor{white}{000}       & \textcolor{white}{000}        \\ 
\cline{1-14}
\grayrow PCA                               &  100    & 100                       & 100     & 100                      &   100  & 100                       & 100     & 100                       &100     &100                       &  99.96      & 100     & 100                       &   9.96   & 9.96     & 9.96         &    20.89   & 20.89     & 20.89       &    6   & 6      & 6       \\
AE                    &    100  & 100                       & 100     & 100                      &    100  & 100                      & 100     & 100                      & 100     & 100                       & 99.91     & 100     & 100                       & 10.14      & 10.14     & 10.14         & 21.58     & 21.58    & 21.58         & 7     & 7     & 7        \\
\hline\hline
\end{tabular}
  \end{adjustbox}
  \vspace{0.5em}
  \caption{Accuracy (ACC), resources and latency for the UNSW and CICIDS use cases, using (S)mall, (M)edium, (L)arge and (H)uge models. Some models are not feasible~(\dag or NF) on Tofino.
  }
  \label{tab:functionality}%
  \vspace{-2em}
\end{table*}%
 
\subsection{Inference Performance}\label{sec:eval-modelfunc}

The inference performance evaluation explores if the mapped in-network ML models have similar inference accuracy as running the same inference task on a server, and how the size of the model affects its accuracy.

\textbf{Results: } The results are presented in Table~\ref{tab:functionality}. As the Accuracy column (left side of the table) shows, for the same model size, all the models have a similar  accuracy performance on the programmable switch as on the sklearn or baseline server, verifying Planter's mapping. 
We evaluate the models in Planter from the following aspects: \textit{i) dataset: } The models on Planter show very similar results to sklearn (for the same model size) on both datasets with different types of attacks. In-network classification results even reach a similar level as on the baseline server with a huge model size. \textit{ii) model type:} different implementations of the same type of model mapping (e.g. DM tree models) show little difference in accuracy. However, the model structure can present different inference capabilities. For instance, NB and KM (LB model) have lower accuracy than other models for the UNSW dataset. \textit{iii) model size: }The right half of Table~\ref{tab:functionality} presents the accuracy performance for different sizes of models. As the model size increases, some models achieve slightly higher accuracy. As larger models require more processing on the switch, we also evaluate the relative memory consumption and the relative latency caused by the in-network inference. The results show that the EB models consume less memory and M/A stages, but lead to longer inference latency than LB models, as the model size increases. In spite of the tolerable impact, such extra overheads can be avoided by deploying the smaller size model and consuming fewer resources, while still achieving a fair accuracy. 

\textbf{switch.p4 Integration: }Planter was integrated with the $switch.p4$, an Intel reference L2/L3 switch, without additional stage consumption. This indicates that standard switching functions can coexist with Planter, with no or minimal cost in stages and latency, but with higher resource utilization.

In general, switch-based solutions have similar results to the same model running on a server using the same selected set of hyper-parameters. How these hyperparameters influence the inference accuracy is discussed next.

\subsection{Scalability}
\label{eva:scalability}
 
This section explores Planter's performance as the model scales up with different hyperparameter settings. UNSW-NB15 dataset is used as the input workload for the evaluation. 


 \subsubsection{Scalability and relative accuracy} Various hyperparameters have a different relative effect of the accuracy performance of an in-network ML model. As an example, we study the effect of action data bits and model depth on models' relative accuracy performance. The number of action bits heavily impacts LB models, while the model depth impacts EB models. The results refer to the relative accuracy, meaning the ratio between the accuracy of switch output and the accuracy of sklearn output.

 \begin{figure}[t]
	\centering
	\begin{minipage}{0.49\linewidth}
	\includegraphics[width=1\linewidth]{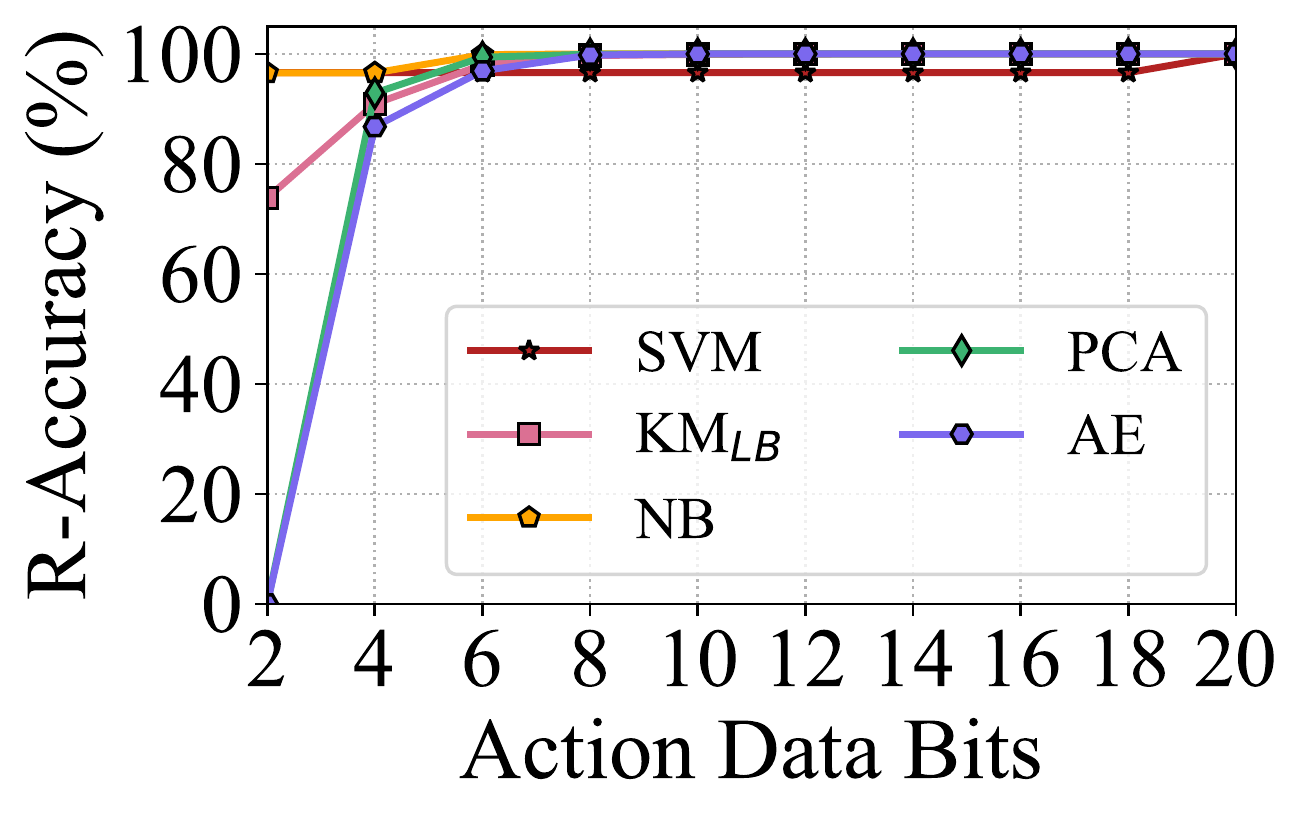}\vspace{-0.5em}\\\centering(a) R-ACC vs Action Bits
\end{minipage}
	\begin{minipage}{0.49\linewidth}
	\includegraphics[width=1\linewidth]{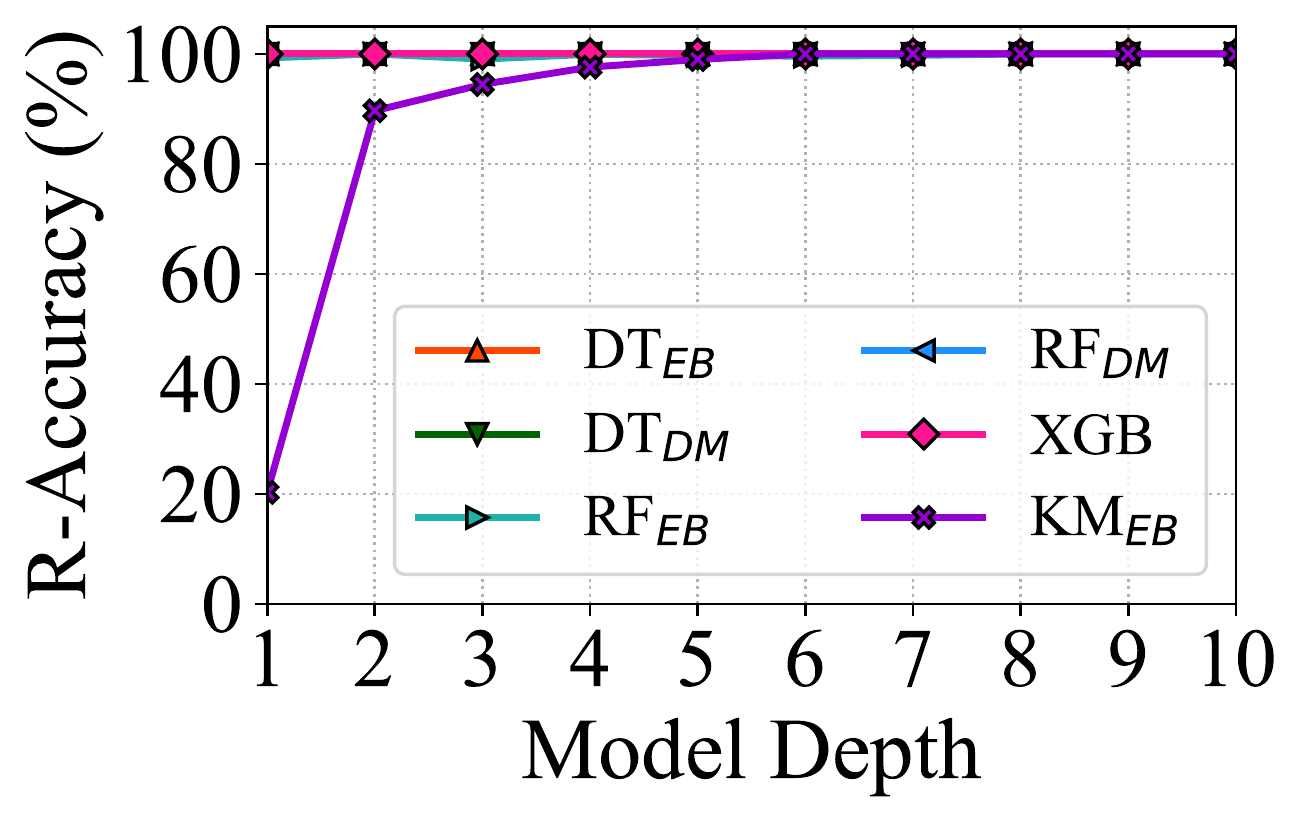}\vspace{-0.5em}\\\centering(b) R-ACC vs Model Depth.
\end{minipage}
	\vspace{-1em}
	\caption{(R)elative-Accuracy between switch and scikit-learn}
	\label{fig:relative acc}
	\vspace{-1em}
\end{figure}

Figure~\ref{fig:relative acc} shows the switch accuracy relative to sklearn accuracy. The relative accuracy increases as action data increase. Except for SVM, the relative accuracy in other models reaches 100\% when 8 action data bits or more are used. The SVM model is more sensitive to the accuracy trade-off of stored intermediate results, and requires 18 action bits to the same accuracy as sklearn.

 \begin{figure}[t]
	\centering
	\begin{minipage}{0.49\linewidth}
	\includegraphics[width=1.02\linewidth]{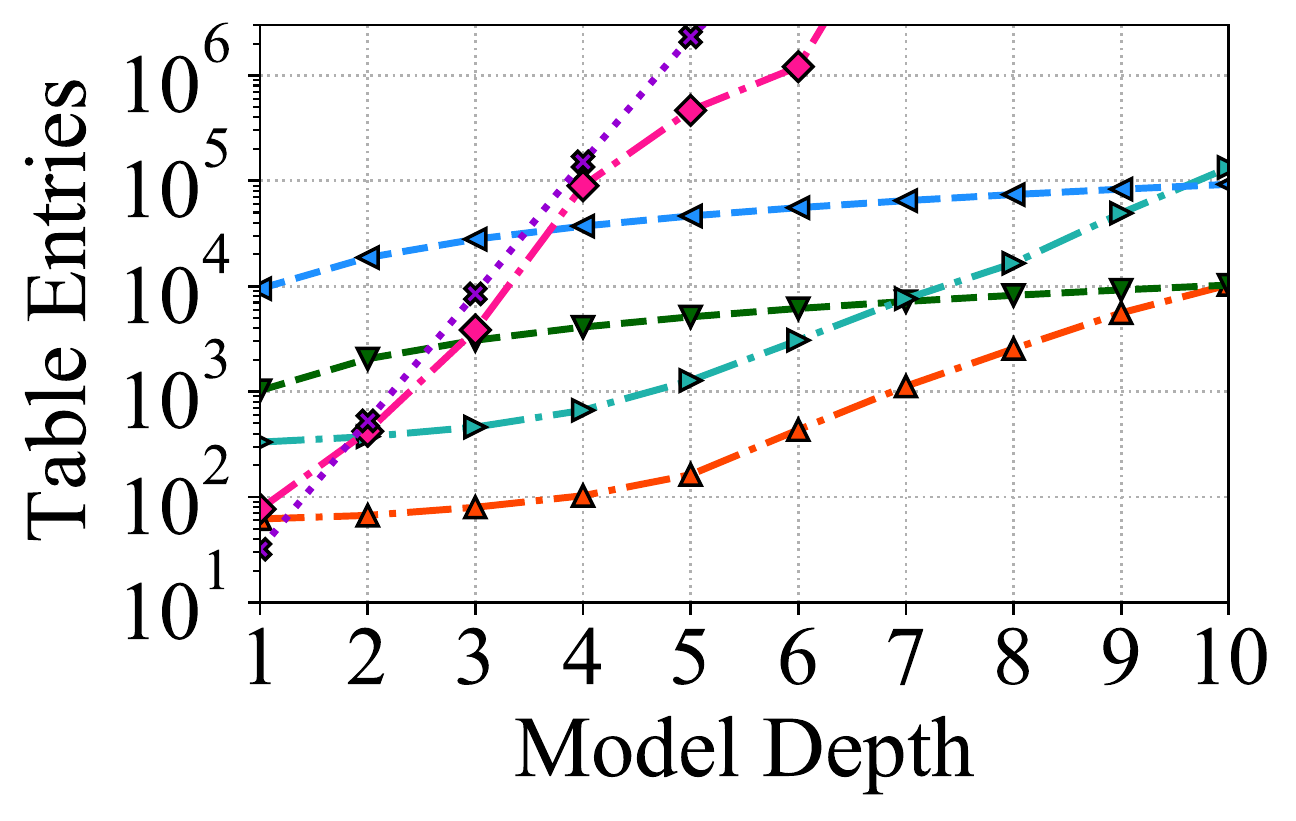}\vspace{-0.5em}\\\centering(a) Model Depth vs Entries
\end{minipage}
	\begin{minipage}{0.49\linewidth}
	\includegraphics[width=1\linewidth]{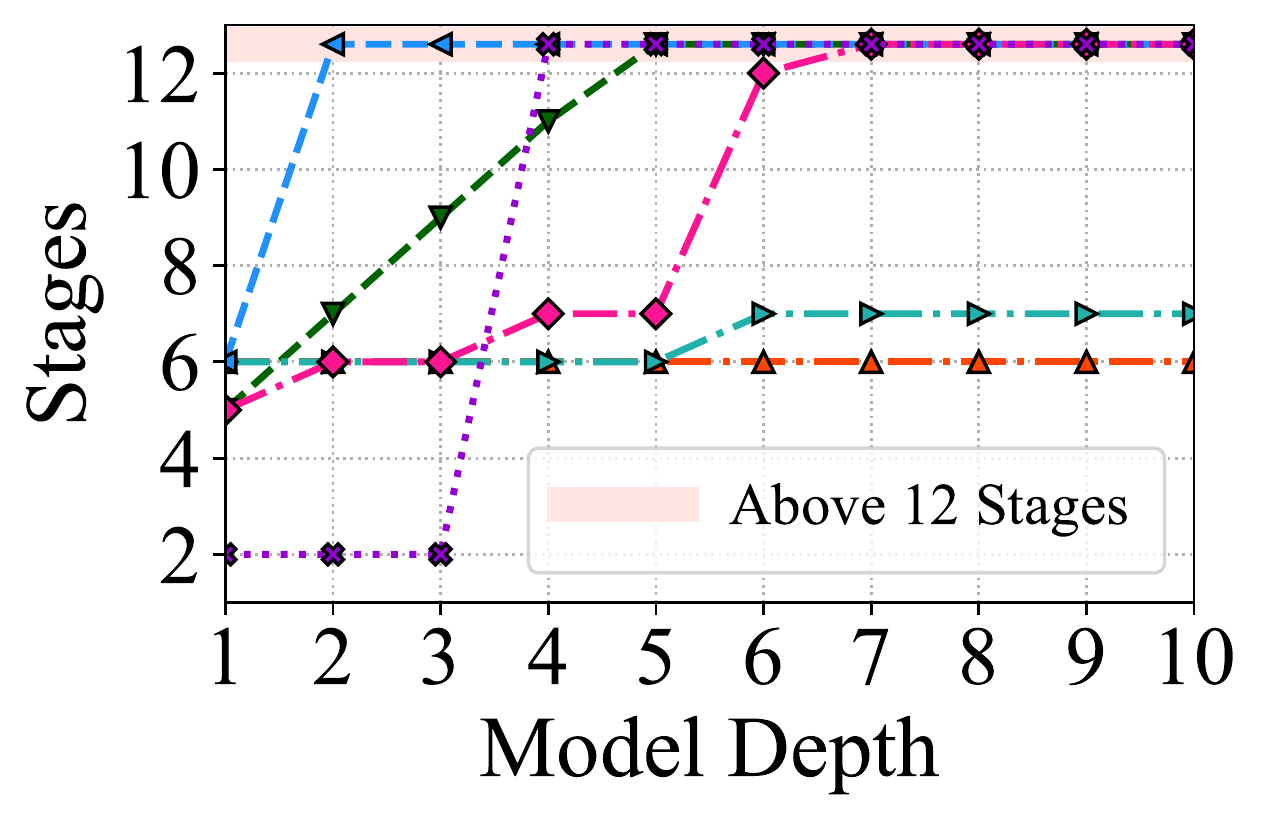}\vspace{-0.5em}\\\centering(b) Model Depth vs Stages.
\end{minipage}
	\vspace{-1em}
\end{figure}

\begin{figure}[t]
\centering
  \begin{minipage}{0.49\linewidth}
	\includegraphics[width=1\linewidth]{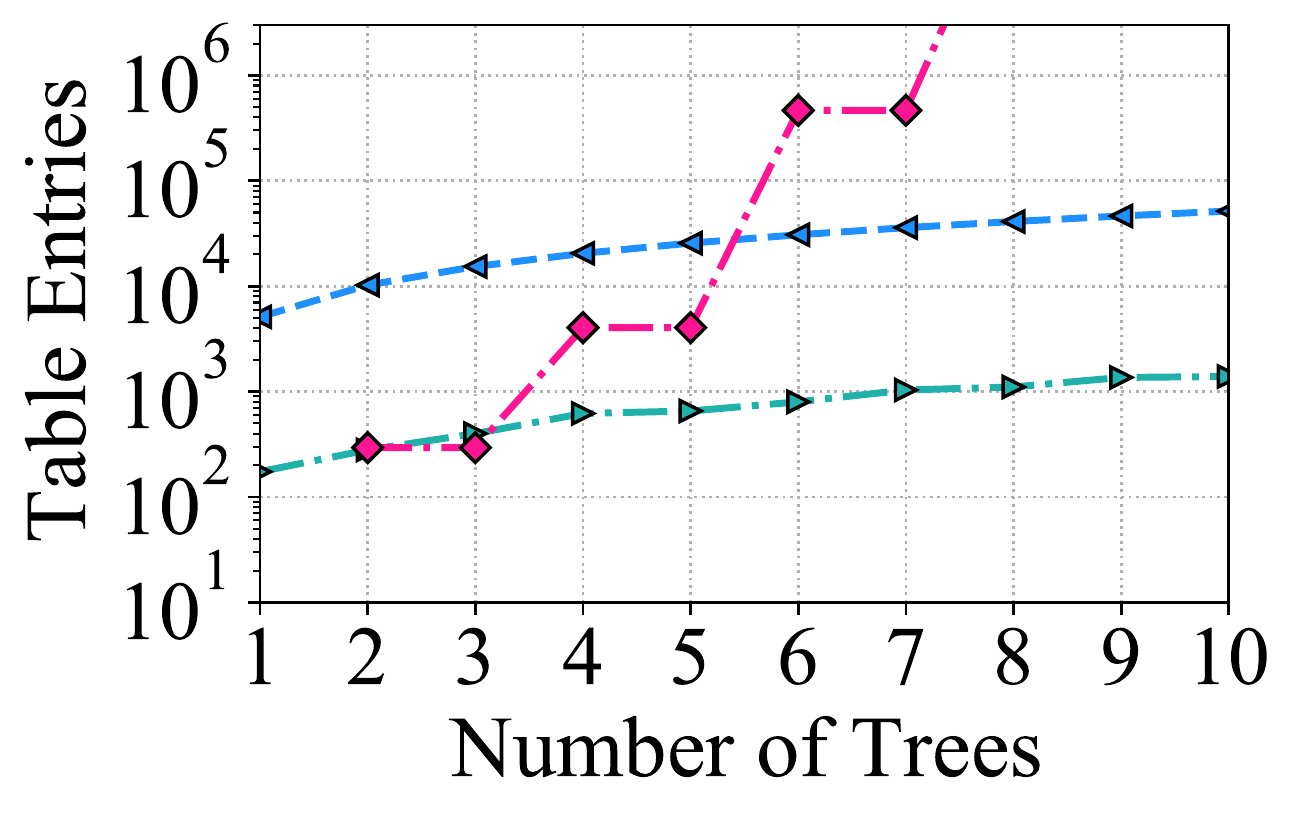}\vspace{-0.5em}\\\centering(c) Number Trees vs Entries
  \end{minipage}
  \begin{minipage}{0.49\linewidth}
	\includegraphics[width=1\linewidth]{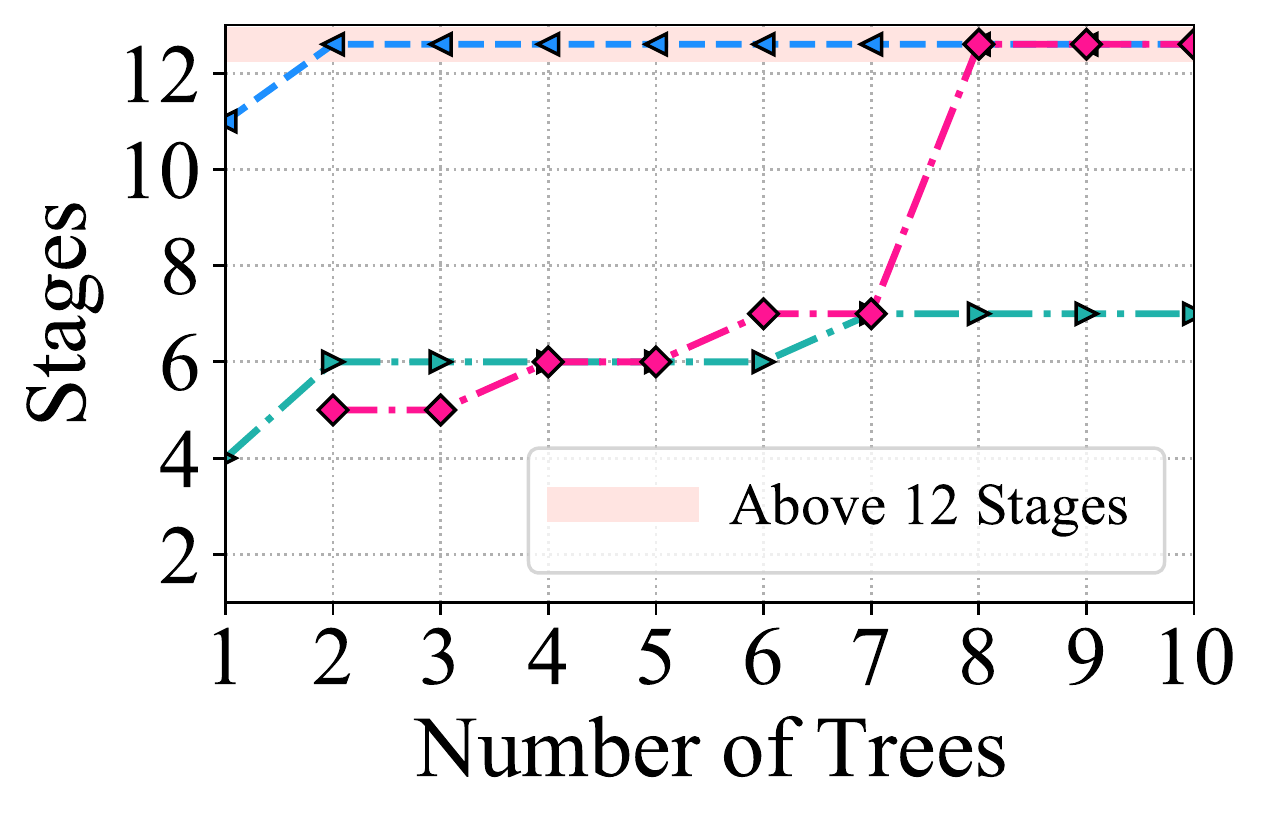}\vspace{-0.5em}\\\centering(d) Number Trees vs Stages.
  \end{minipage}
\vspace{-1em}
\end{figure}

 \begin{figure}[t]
	\centering
	\begin{minipage}{0.49\linewidth}
	\includegraphics[width=1\linewidth]{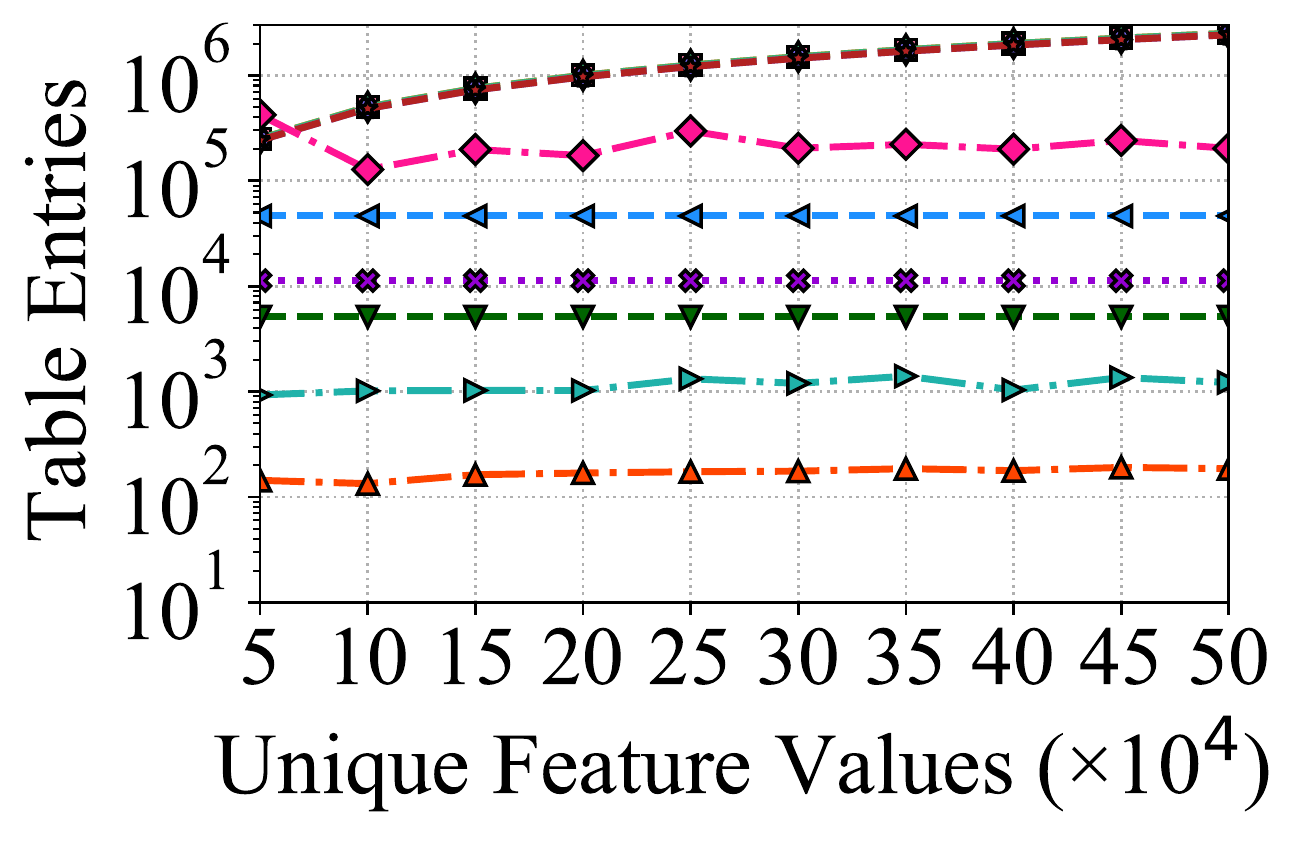}\vspace{-0.5em}\\\centering(e) Unique Values vs Entries
\end{minipage}
	\begin{minipage}{0.49\linewidth}
	\includegraphics[width=1\linewidth]{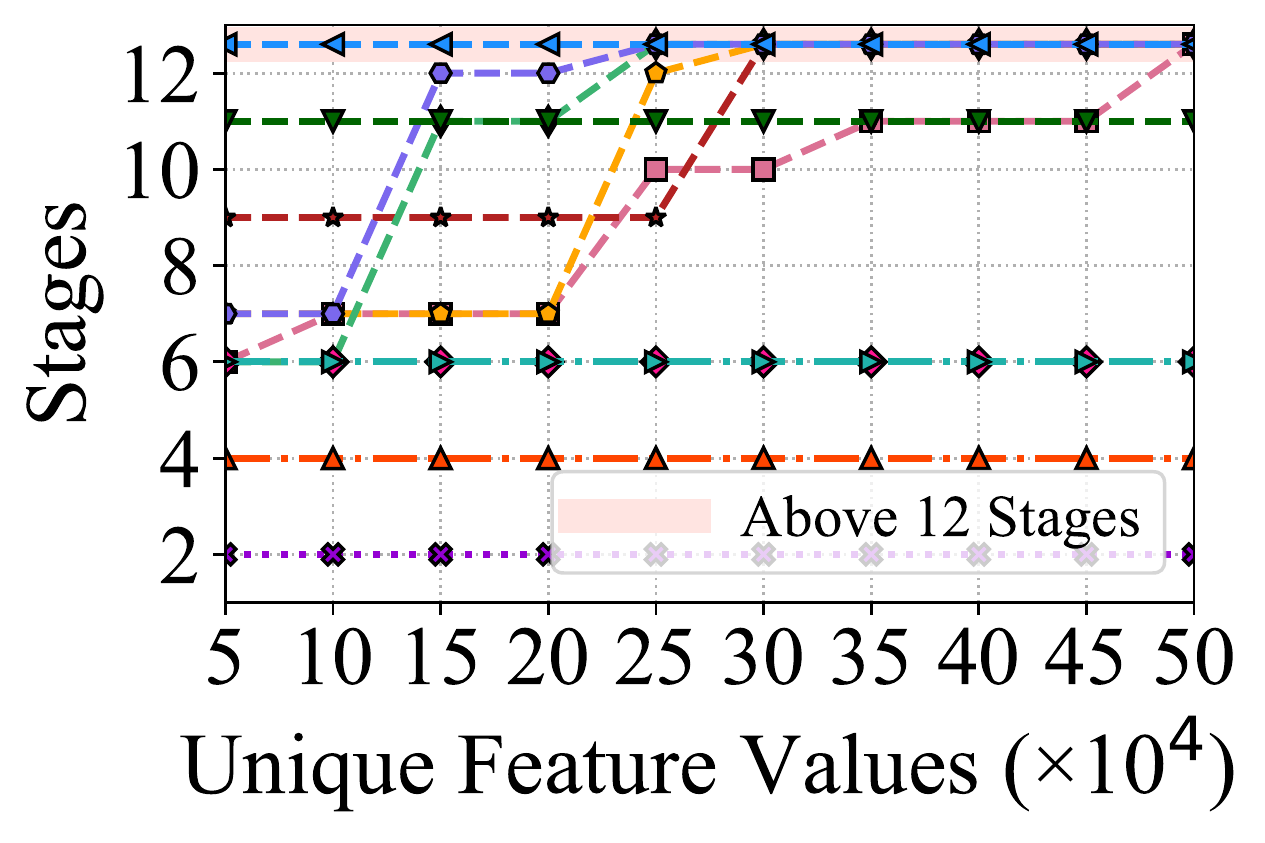}\vspace{-0.5em}\\\centering(f) Unique Values vs Stages.
\end{minipage}
	\vspace{-1em}
\end{figure}

\begin{figure}[t]
\centering
  \begin{minipage}{0.49\linewidth}
	\includegraphics[width=1\linewidth]{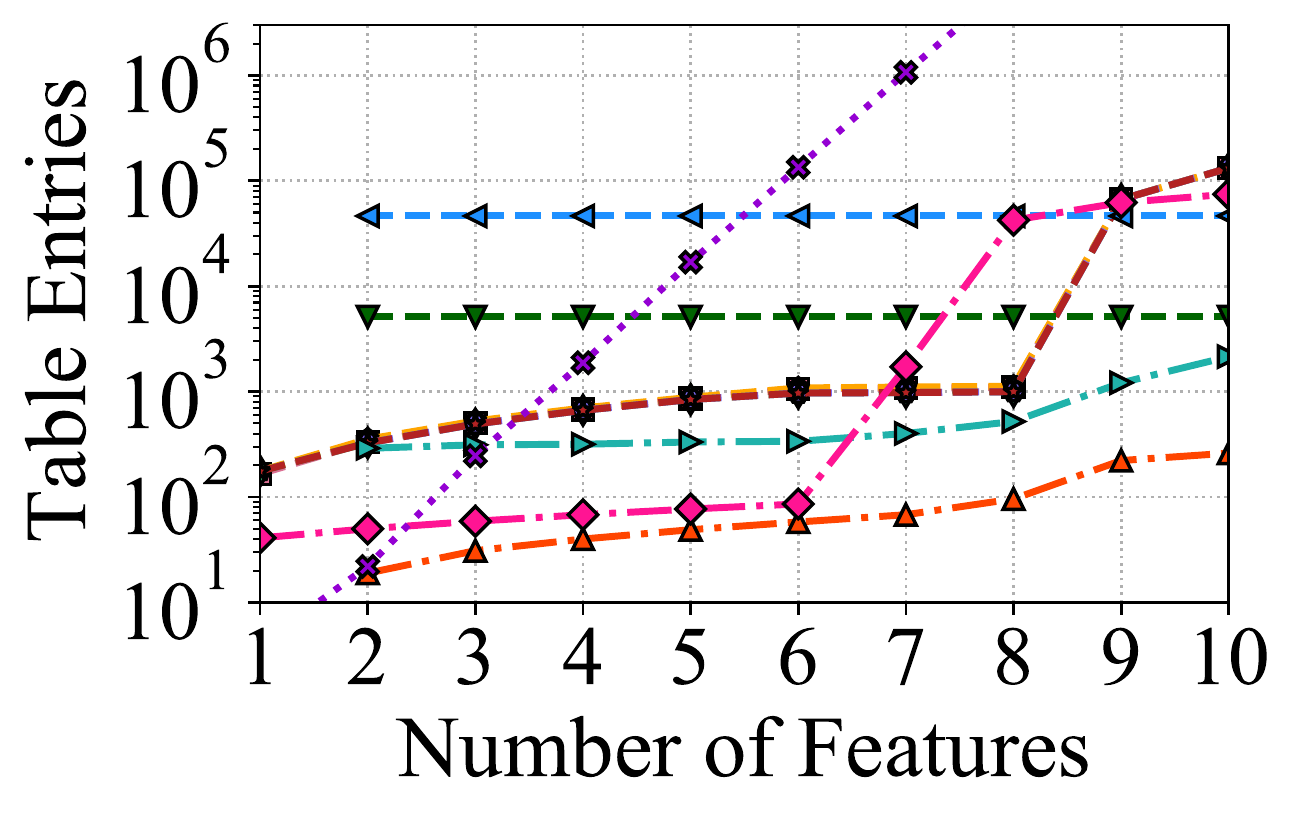}\vspace{-0.5em}\\\centering(g) Num Features vs Entries
  \end{minipage}
  \begin{minipage}{0.49\linewidth}
	\includegraphics[width=1\linewidth]{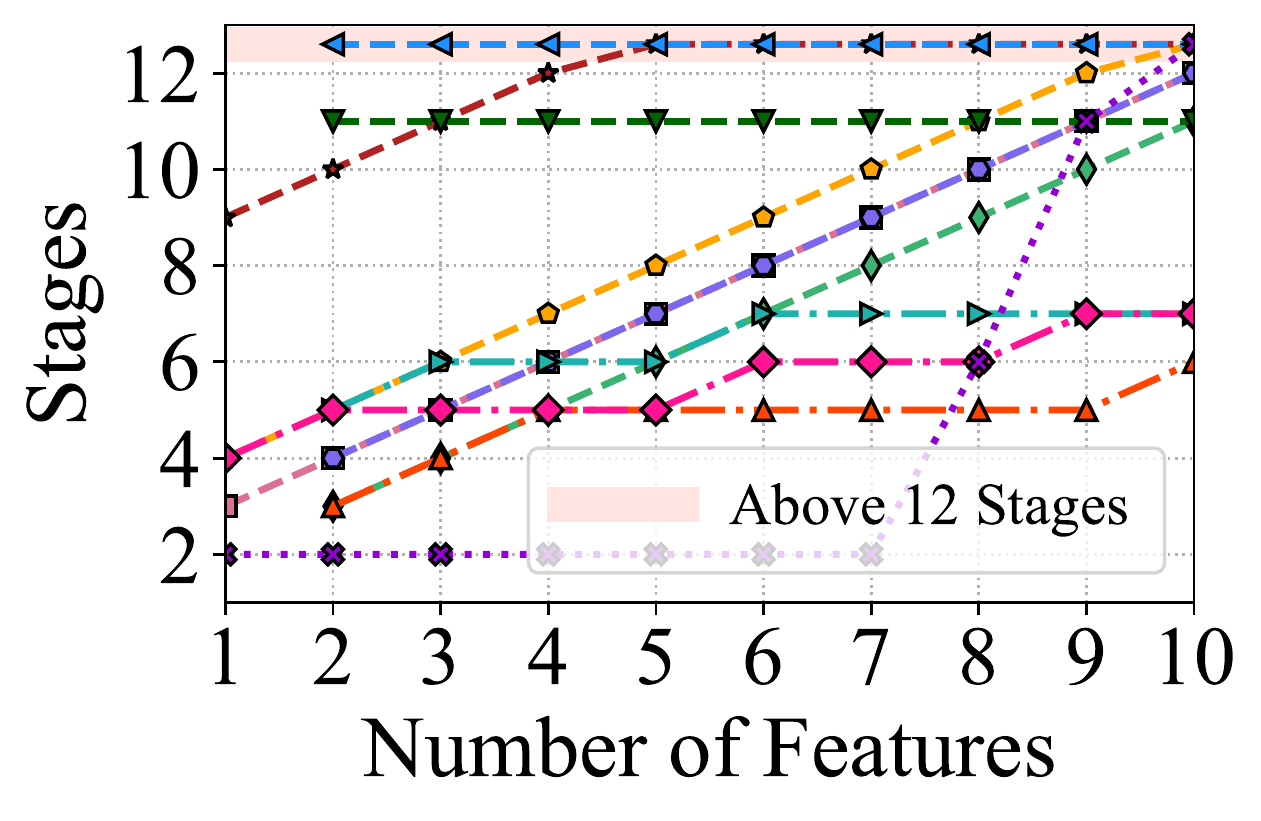}\vspace{-0.5em}\\\centering(h) Num Features vs Stages.
  \end{minipage}
\vspace{-0.5em}
\includegraphics[width=1\linewidth]{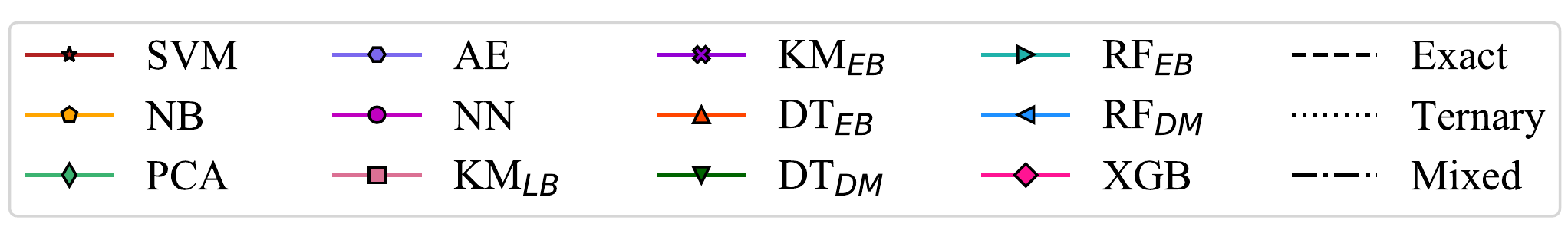}
	\vspace{-1em}
	\caption{Memory and stage scaling with the model hyperparameters and feature properties.}\label{fig:scal1}
	\vspace{-1em}
\end{figure}
\subsubsection{Resources scalability}

Resource scalability evaluation focuses on the number of table entries and the number of pipeline stages. Table entries indicate the potential memory requirement from the switch, and the number of stages remaining M/A stages for model growth and non-parallel functionality. Two dimensions are evaluated: model/convert hyperparameters (model depth, action data bits, and number of trees) and dataset inputs (number of features and unique feature values). 

\begin{figure}[t]
\centering
  \begin{minipage}{0.49\linewidth}
	\includegraphics[width=1\linewidth]{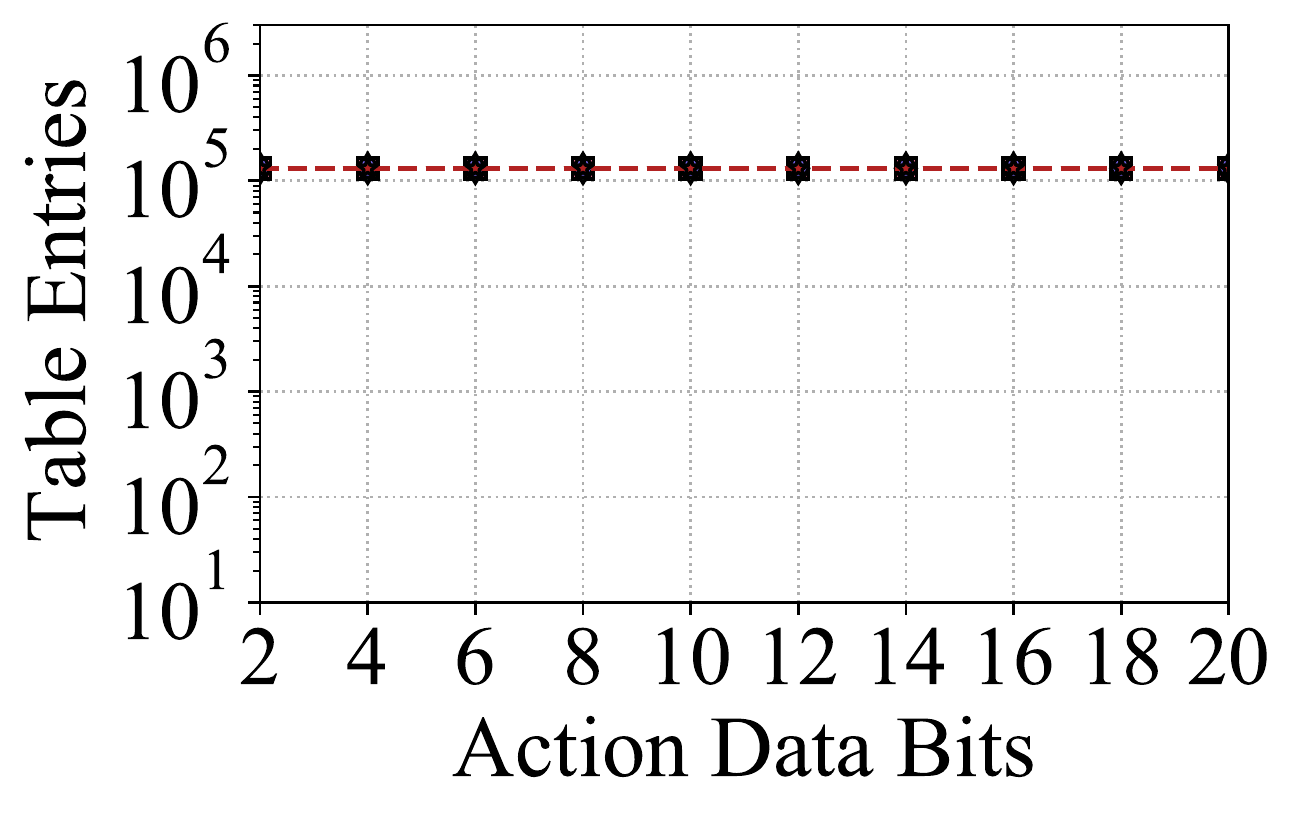}\vspace{-0.5em}\\\centering(a) Action Bits vs Entries
  \end{minipage}
  \begin{minipage}{0.49\linewidth}
	\includegraphics[width=1\linewidth]{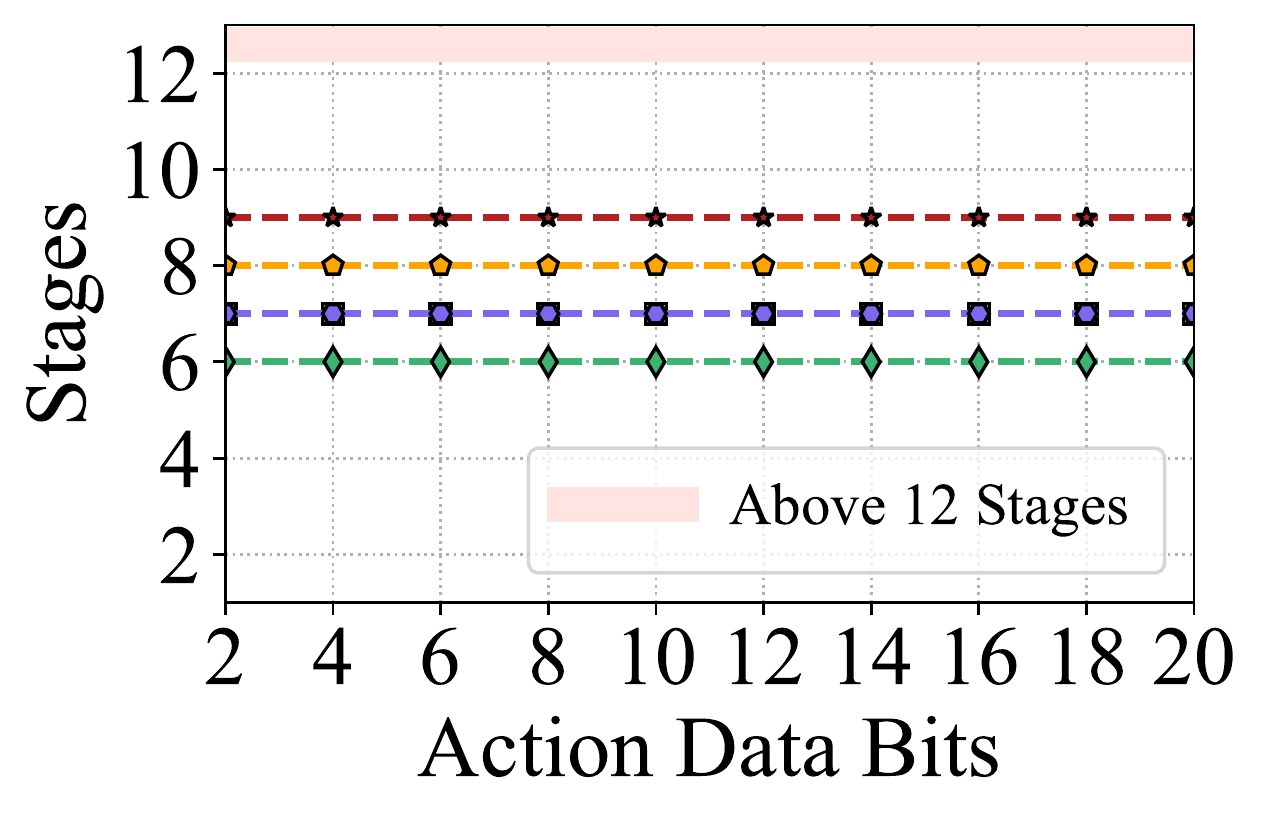}\vspace{-0.5em}\\\centering(b) Action Bits vs Stages.
  \end{minipage}
\vspace{-0.5em}
\includegraphics[width=1\linewidth]{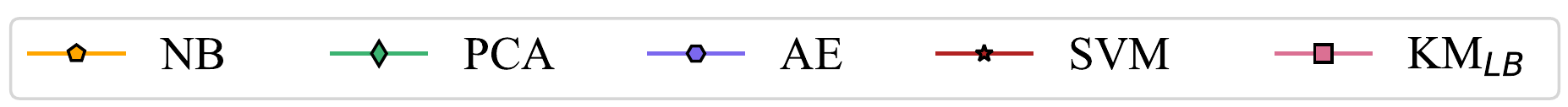}
	\vspace{-1.5em}
	\caption{Action bits effect on LB solutions scalability.}\label{fig:scal2}
	\vspace{-1em}
\end{figure}

\textbf{Results: } Figure \ref{fig:scal1} (a) \& (b) show that as a model's depth increases, more table entries are required in all EB models and direct-mapping tree-based models. Among them, DM solutions have a comparatively slower increment. EB tree models are more stable in terms of stage consumption. 
Figure \ref{fig:scal1} (c) \& (d) show that as the number of trees increases,  EB tree models require 8 stages less than DM tree models, unless the number of table entries is excessive.
In Figure \ref{fig:scal1} (e) \& (f), the feature range, which is the number of unique feature values per feature, only influences LB models' stage and memory consumption.
Figure \ref{fig:scal1} (g) \& (h) show that except for DM tree-based models, models consume more table entries as the number of  feature values increases. In terms of stage consumption, only LB based models have a strong correlation to the number of features.
In Figure \ref{fig:scal2} (a) \& (b), show that the number of action data bits does not influence the required number of table entries and the required number of stages. Note that the evaluated models are those where action bits are a parameter. In others, the other of action of action bits is a result of other hyperparameters (e.g., depth).

The insights from this evaluation are:
\begin{enumerate} 
    \item LB-based mapped models are sensitive to use case characteristics more than model/convert hyperparameters. EB models have a relatively steady number of stages, but their scalability is influenced by the number of required entries (e.g., range of feature values). DM approach has the best scalability in table entries requirements, but performs badly in terms of stages consumption.
    \item When the size of a model is changed (S/M/L), EB tree models have advantage in controlling the number of stage compared with DM tree models. In contrast to KM$_{LB}$, KM$_{EB}$ has a more steep trend in the consumption of table entries but uses less stages when the model is small.
    \item Table entries and stage consumption are determined mainly by the model mapping methodology. Under extreme circumstances, too many table entries can increase the number of M/A stages used. For example, in Figure \ref{fig:scal1} (c) \& (d), due to the number of table entries,  XGB has a different trend compared to RF$_{EB}$, when the number of trees exceeds 7.
\end{enumerate}

\subsection{Baselines and Comparisons}
This section evaluates the merit of the new encoded-based (EB) tree models design and the upgraded NB model, in terms of resources.

 \begin{figure}[htbp]
	\centering
	\begin{minipage}{0.49\linewidth}
	\includegraphics[width=1\linewidth]{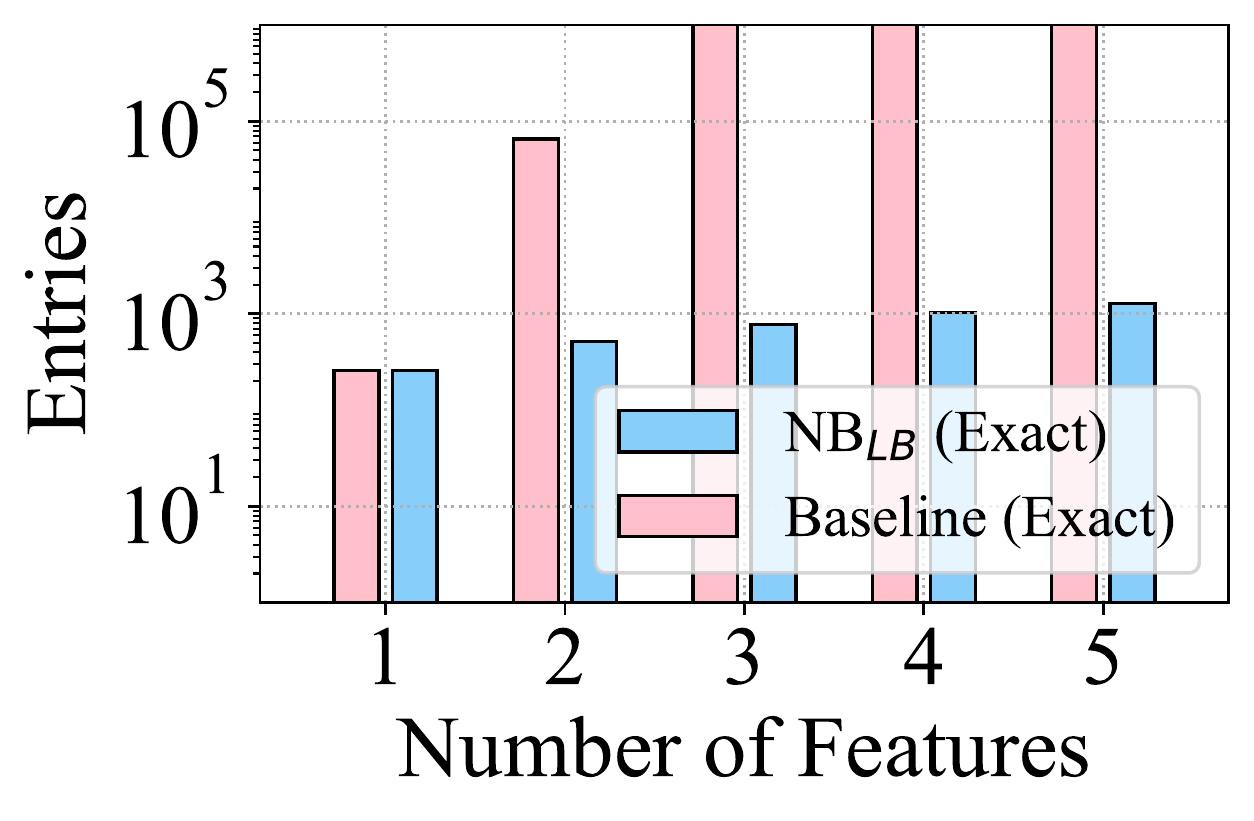}\vspace{-0.5em}\\\centering(a) NB$_{LB}$ vs NB Baseline
\end{minipage}
	\begin{minipage}{0.49\linewidth}
	\includegraphics[width=1\linewidth]{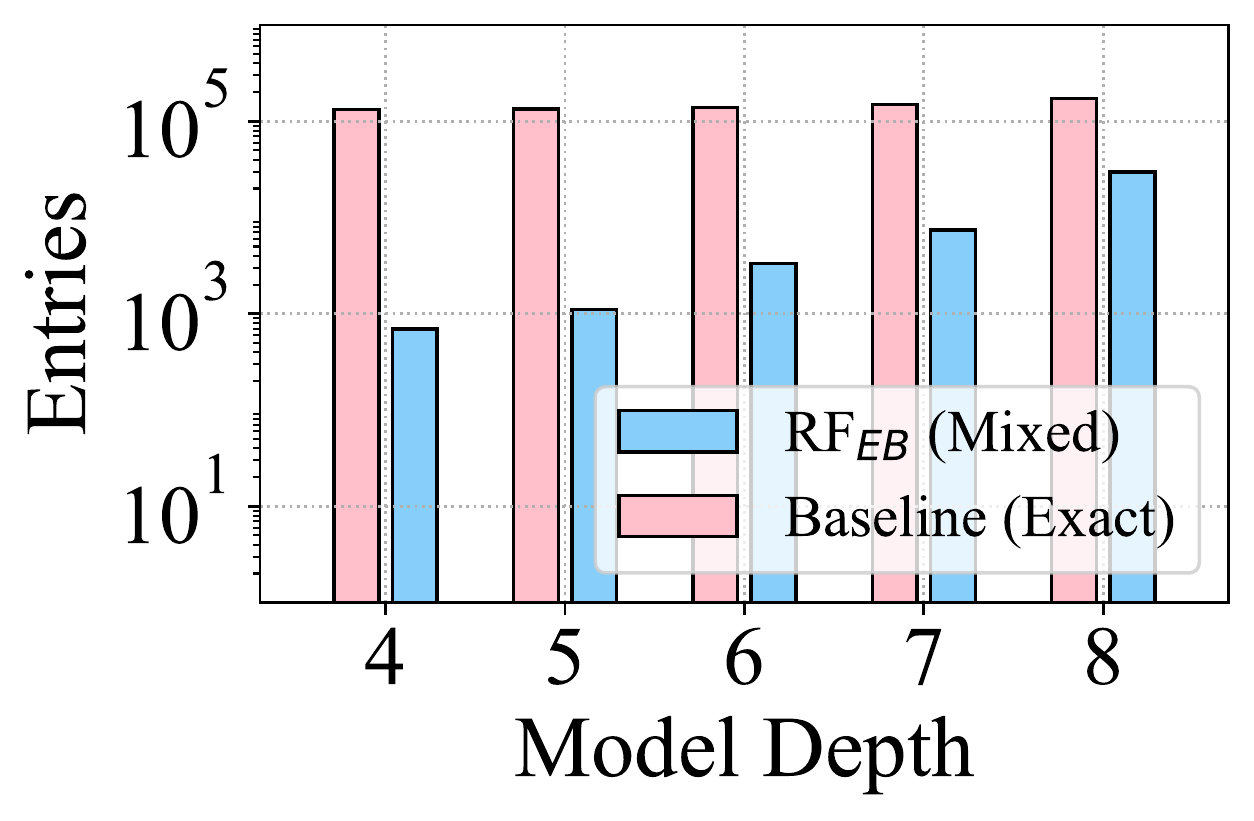}\vspace{-0.5em}\\\centering(b) RF$_{EB}$ vs RF Baseline.
\end{minipage}
	\vspace{-0.7em}
	\caption{Table entries in upgraded models compared with the baseline implementation.}
\label{fig:compare to baseline}
	\vspace{-0.5em}
\end{figure}

EB tree models are compared with the baseline from~\cite{xiong2019switches}, which used only exact match tables and no default actions. As shown in Figure \ref{fig:scal1}, with the help of M/A stage sharing, EB tree models perform much better in terms of growth in stage number. In addition, as shown in Figure \ref{fig:compare to baseline} (a), the upgraded NB requires fewer entries compared to IIsy's NB (when multiplication operation is not allowed). 
Except for Clustreams (KM$_{EB}$), most of the in-network ML algorithms use only exact match. Widely used ternary or LPM tables can significantly reduce table entries usage. As shown in Figure~\ref{fig:compare to baseline} (b), take RF$_{EB}$ as an example, Planter's EB tree model variations use less table entries compared with the baseline. Clustreams performs well when the number of features is small, and the range of unique feature values is large, otherwise it is outperformed by KB$_{LB}$.

\subsection{Throughput \& Latency}
The evaluation of throughput of different models is shown under the attack detection use case, which is a volumetric use case. Latency is shown using financial transaction prediction use case, which is latency sensitive. Setup details are provided in Appendix \ref{Use case Setups}. The results presented in this section are a subset of the tests.

\begin{figure}[htbp]
	\centering
	\includegraphics[width=1\columnwidth]{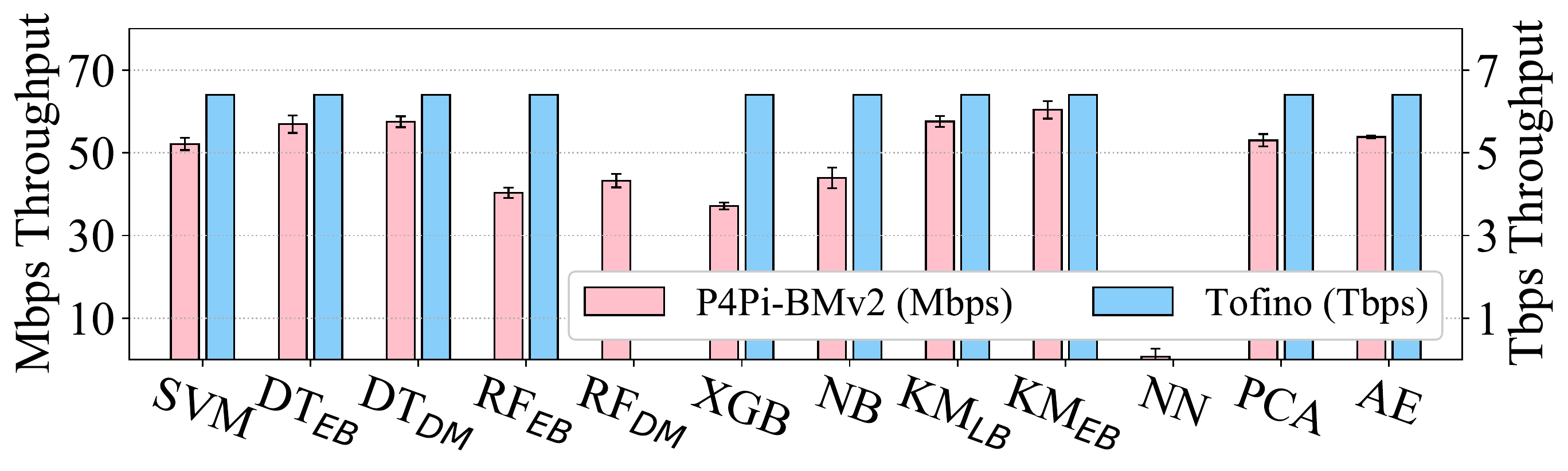}
    \vspace{-2.5em}
	\caption{Throughput of ML algorithms for attack detection on Tofino (in Tbps) and P4Pi (in Mbps).}
    \label{fig:eval-throughput}
	\vspace{-1em}
\end{figure}

Throughput tests record the throughput of each in-network ML algorithm on a Tofino switch and P4Pi, as shown in Fig.~\ref{fig:eval-throughput}. The baseline throughput of basic forwarding is 6.4 Tbps on Tofino and 64 Mbps on P4Pi. On a Tofino switch, full 6.4Tbps is achieved for all feasible models (Table~\ref{tab:functionality}. On P4Pi, the results vary for different models. Seven of the models achieve more than 80\% of the baseline throughput. Ensemble models (RF$_{EB}$, RF$_{DM}$, and XGB) and NN have degraded throughput on P4Pi, due to their increased use of resources. XGB and RF$_{EB}$ do run at line rate on Tofino.


\begin{figure}[htbp]
	\includegraphics[width=1\columnwidth]{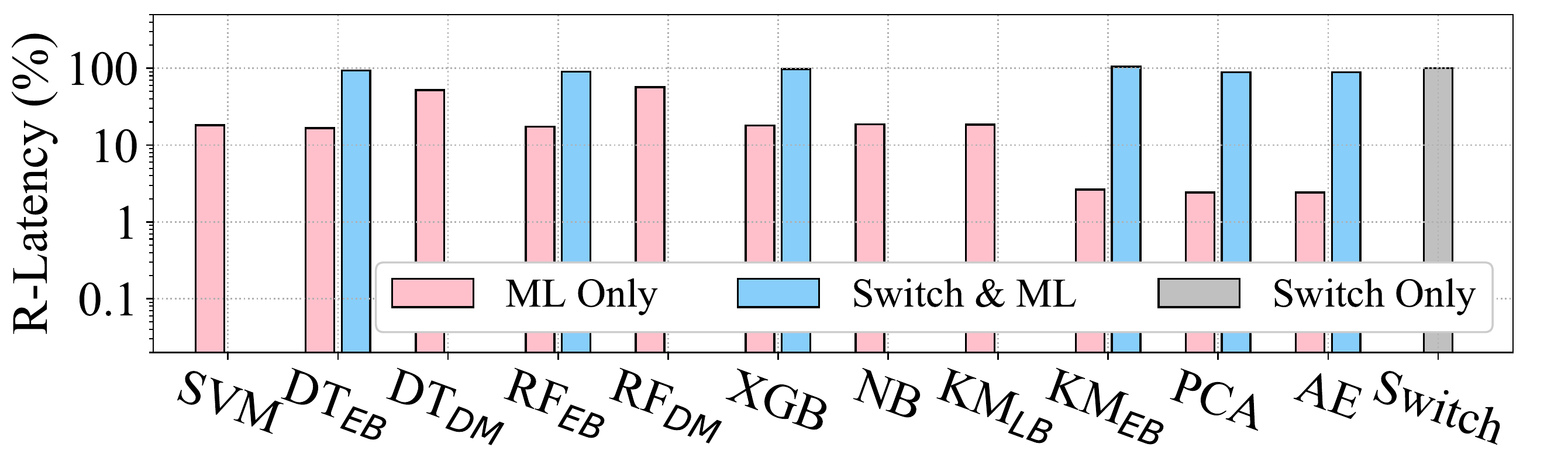}
    \vspace{-2.5em}
	\caption{The relative latency (R-Latency) on Tofino in the financial prediction use case, measured for standalone ML, ML combined with $switch.p4$, and standalone $switch.p4$.}
    \label{fig:eval-latency}
	\vspace{-1em}
\end{figure}

In latency tests, relative latency results based on Tofino are illustrated in Figure \ref{fig:eval-latency}. The baseline is the latency of \textit{switch.p4}. 
When only ML models are deployed, without additional functions, the latency is lower than 22\% of switch.p4 in most cases. When the ML models are combined with switch.p4, there is an overhead of less than $4.7\%$ for all applicable algorithms. EB solutions and LB dimensional reduction algorithms have better compatibility with other switch functions in resource-constrained targets, as less logic is required.







\section{Discussion}
 
\paragraph{ML Performance} Planter-based ML models provide inference accuracy similar to running the exact same model on the host, as the evaluation shows. However,  there is a trade-off between model size and inference accuracy. In some cases, a large model can achieve higher accuracy at the price of additional switch resources. While Planter can't avoid that, the models that fit on a switch, while coexisting with standard functionality, achieve high accuracy. 

\paragraph{Pipeline Stages} The number of stages required by a model relates both to the type of mapped model and its size. For the UNSW dataset, at least 2 stages are consumed, and some models do not fit. This is where the benefits of encode-based solutions are shown over direct-mapping solutions.
Planter also shows that many of the stages can be shared with standard switch functionality. Moreover, some designs can be hand-modified to reduce stages, e.g., where network and ML functions have similarities. Our experience shows that a manually optimized code can save 2-3 stages, by improving the final logic and forcing stage allocation.
%

\paragraph{Use cases} Planter provides a one-click in-network ML solution for emerging use cases. Due to space limitations, only a subset of datasets and use cases are presented, including in Appendix~\ref{apdix: eval-add-dataset}. While current use-cases of in-network ML are focused on network applications and network security, we believe this is a chicken-and-egg problem, due to the lack of a suitable framework. Planter aims to be to in-network ML what CUDA was to GPUs~\cite{greatcuda} the enabler for wide adoption, leading to a proliferation of use cases.

\paragraph{Future Work} Extensions of the Planter framework focus on targets and resource consumption. This includes adding support for a smartNIC or FPGA, along with the required architecture support. From resource perspective, this includes minimizing table sizes, which is also expected to reduce runtime. 



\section{Conclusion}\label{ch12-Conclusion}


This paper presented Planter, a modular framework for one-click implementation of in-network ML algorithms. Planter's modular design enables integration of new ML models, architectures, targets, and use cases. 
Planter implements a wide range of in-network ML algorithms, including two new dimensional reduction algorithms, and an upgrade to two previously proposed algorithms.  
The evaluation shows that Planter provides accurate mapping of trained models to a switch, can achieve high accuracy and line rate throughput, and can be integrated with switch.p4 without consuming additional stages.
As an open-source platform, Planter is the enabler for the research of in-network machine learning, and its code is available at~\cite{planterrepo}.

This paper complies with all applicable ethical standards of the authors' home institutiona.\\

\noindent\textbf{Acknowledgements}\label{sec:acknowledgement}
This work was partly funded by VM-Ware. We acknowledge support from Intel.


\bibliographystyle{acm}
\bibliography{sigproc} 

\appendix

\section{List of Acronyms}\label{sec:app_acronym}

The following acronyms, as shown in Figure \ref{table: notations}, are used in this paper.
\begin{table}[h]
\begin{adjustbox}{width=1\linewidth,center}
  \centering 
    \begin{tabular}{c|l}
    \hline
    Acronyms    & Definition \\ 
    \hline
    \grayrow $AE$         & Autoencoder\\
     $BNN$     & Binary Neural Network \\
    \grayrow  $DM$     & Direct Mapping  \\
    $DT$         & Decision Tree\\
    \grayrow $DT_{DM}$         & Direct Mapping Decision Tree (SwitchTree, pForest)\\
      $DT_{EB}$     & Encode-based Decision Tree (IIsy)\\
     \grayrow  $EB$     & Encode-based  \\
     $FP$ & False Positive    \\
    \grayrow $ FN$  &   False  Negative   \\
    $HFT$  & High Frequency Trading     \\
    \grayrow $KM$         & K-means\\
     $KM_{EB}$     & Encode-based K-means (Clustreams)\\
    \grayrow $KM_{LB}$         & Lookup-based K-means (IIsy)\\
    $LB$     &  Lookup-based \\
    \grayrow $ LOC $     & Lines of Codes  \\
     $ML$     &  Machine Learning \\
    \grayrow $MS$     & Multiple Solutions  \\
     $M/A$     &  Match/Action \\
    \grayrow  $NB$     & Naïve Bayes \\
     $ NDA $  & Non-Disclosure Agreement    \\
    \grayrow $ NF$  &   Not Feasible     \\
     $PC$     & Peer Comparison  \\
    \grayrow $PCA$         & Principal Component Analysis\\
     $PISA$     & Protocol Independent Switch Architecture  \\
    \grayrow $PP$     & Plug-and-Play ability enabled  \\
     $PSA$  & Portable Switch Architecture   \\
    \grayrow $P4$     &  Protocol-Independent Packet Processors \\
     $PNA$     &  Portable NIC Architecture \\
    \grayrow $QoE$     & Quality of Experience  \\
     $RF$     & Random Forest \\
    \grayrow $RF_{DM}$     & Direct Mapping Random Forest (SwitchTree, pForest)\\
     $RF_{EB}$         & Encode-based Random Forest (Planter)\\
    \grayrow $R-ACC$     & Relative Accuracy \\
     $R-Latency$     & Relative Latency \\
    \grayrow $SC$     & Source Code available  \\
     $SVM$     & Support Vector Machine \\
    \grayrow $S/M/L/H$     &  Small/Medium/Large/Huge \\
     $ TN$  &   True Negative      \\
    \grayrow $TNA$     & Tofino Native Architecture \\
     $TP$     &  True Positive \\
    \grayrow $XGB$         & Extreme Gradient Boosting (XGBoost)\\
      \hline
     \end{tabular}%
  \end{adjustbox}
  \vspace{0.3em}
\caption{Acronyms.}
\label{table: notations}
\end{table}

\section{Modular Design Details}
\label{ch3-Planter-modular}

Planter is a modular framework. Each module is in charge of their own function without influences. Many ML model modules, Architecture Models, Target Modules, and Use Case modules exist in the framework. Based on actual needs, user can design their own module and plug it into other modules or simply select existing modules. There are usually four modular designed components: Modular Models, Modular Architectures, Modular Targets, and Modular Use Cases.\\
\begin{itemize}

\item Models Modularity. Each model-related script is stored by two files in a folder. All these folders are stored under $./src/models/$. Main file $main.py$ integrates Model Trainer and Model Converter and file $dedicate\_p4.py$ focus on generating Dedicate P4 codes. These codes include Metadata and part of Ingress and Egress pipeline.\\

\item Architectures Modularity. Architecture defines the st-ructure of the P4 codes. Like a shell of the hermit crab, use case and model-related P4 codes live in it. In the Planter framework, There are parallel architecture folders under $./src/architectures/$ that support different architectures. Under each folder, there is a $standard\_p4.py$ file to generate architecture-specific P4 codes, for example, $includes$ and pipeline structures. The Planter main program chooses the selected architecture module based on configurations.\\

\item Targets Modularity. Modular targets allow new designs to be easily compiled, tested, and deployed. Each target uses two python scripts: Model Compiler $run\_model.py$ and Model Tester $test\_model.py$ in a Target folder under directory $./src/targets/$. The model compiler compiles the generated P4 codes and prepares the switch model. The model tester will send packets to the switch model for testing.\\

\item Use Cases Modularity. \label{modular_use_case} The same model and architecture may support different use cases. One use case may also be used on different data plane devices. Modular use case hides under architecture folder. Under $./src/architectures/$, each use case has a folder. Under the folder, there is a Python file $common\_p4.py$ file, which is responsible for generating use case related headers, parsers, and metadata. \\

\end{itemize}

\noindent Besides, under $./src/funtions/$, Planter supports many useful functions, such as exact-to-ternary and exact-to-LPM table transformer. These functions can assist in-network ML and in-network computing algorithms. 

\section{Use case Setups}
\label{Use case Setups}
Considering the throughput requirement, we conduct the throughput tests under attack detection use case to evaluate the potential overhead from Planter. To achieve the attack detection function, we use the UNSW dataset for training and 5-tuple traffic features as feature sets. We enable the basic forwarding function as in \textit{p4lang/tutorials} as the baseline. In parallel, we deploy attack detection function with ML models in Planter. To configure the Planter to support the use case function, we modify the \textit{common\_p4.py} file for L3/L4 protocol header and parser definitions. The 5-tuple features are extracted to metadata once the header fields are parsed. The features stored in metadata are then input and processed to the converted ML model. A detection decision is then output, deciding if the packet can be forwarded or dropped.  

For the financial prediction use case, we aim to predict future stock price movements based on the historical record of trade transaction data files provided by NASDAQ. Given the raw data feed, we reconstruct a csv file containing order messages using an open-source constructor \cite{https://doi.org/10.5281/zenodo.5209267}. Based on the messages adding a new order, we use the side (whether an order is buy or sell), size, and price of individual message as features. Labels are created based on the change of mid-price which can indicate price movements. It is worth noting that the \textit{common\_p4.py} file can be modified for a customized protocol header (Nasdaq especially uses ITCH protocol to communicate market data \cite{jepsen2020forwarding}) and high-level feature extraction. Additionally, in the Jane Street Market Prediction dataset, to make the case closer to reality, for each trading opportunity, Planter allows each feature data to be encapsulated inside a specialized protocol or in the payload with ASCII format (csv in payloads).  

\section{Testbed Description}\label{sec:app_setup}
The system test environment uses APS-Networks BF6064X, an Intel Barefoot Tofino platform with $64\times100G$ ports. 
The switch runs Ubuntu 18.04.1 and Barefoot's SDE 9.6.0 is used on the switch. The software development environment uses SDE 9.4.0.

ESC4000A-E10 servers using AMD EPYC 7302P  CPUs with 256GB RAM, Ubuntu 20.04LTS, and equipped with Mellanox ConnectX-5 100G NICs are used to send traffic to the switch using DPDK 20.11.1 and PktGen 21.03.0. Four CPU cores are dedicated per port. 

To test full throughput, a snake configuration is used, where traffic is looped from each port to the following one, enabling traffic across all 64 ports, which is a common practice~\cite{dang2020p4xos}. A set of python scripts is used to generate, capture and check traffic. Simple forwarding achieves on this baseline 6.4Tbps on the switch.

The P4Pi environment uses Raspberry Pi 4 Model B with 8GB of RAM. It runs P4Pi code released v0.0.3.
The throughput test is conducted referring to the benchmark python script for BMv2 performance test with performance mode configured as suggested in \cite{p4performance}.  
The Raspberry Pi set as P4Pi running v1model over BMv2 software switch.

\section{Evaluation Details}
\label{Evaluation Details}

\subsection{Evaluation Metrics: }
For each part of evaluation, the detailed explanation of our used metrics is as follows: 

\begin{itemize}
    \item[(1)] \textit{Accuracy}. $ACC = \frac{TP+TN}{TP+TN+FP+FN}$, shows the percentage of correct inference. It directly reflects many percentage of data points are correctly classified.
    \item[(2)] \textit{F1 score}. $F1 = \frac{2TP}{2TP+FP+FN}$, shows a more comprehensive inference performance of each class. We use the macro result to avoid the biased label distribution misleading the classification result.
    \item[(3)] \textit{Pearson correlation coefficient}. $\rho_{X, Y}=\frac{\operatorname{cov}(X, Y)}{\sigma_{X} \sigma_{Y}}$, is used for all dimensional reduction algorithms to represent the linear correlation between two sets of data. For the system performance portion, following the NDA, we record the relative memory and latency of each model to the reference switch program: $switch.p4$.
    \item[(4)] \textit{Memory \& Table Entries}. The memory consumption of each in-network ML model directly depends on the total number of generated table entries. Except for the memory on Tofino, we use the number of table entries to refer to memory consumption.
    \item[(5)] \textit{Latency \& Stage}. The latency on each in-network ML model is highly correlated to the number of stages consumed. The stage is the key index depending on whether the mapped ML algorithms can be deployed on the network devices. The latency and stage consumption we record depend on the ML model itself. As the models can be executed in parallel with the networking functions, they are not equal to the extra costs.
     \item[(6)] \textit{Throughput}. We record the switch throughput as the average value among 5 independent trial results. 
\end{itemize}
\begin{table}[htbp]
    \begin{adjustbox}{width=1\linewidth,center}

\centering 
    \begin{tabular}{cccccccc}
    \hline\hline
    \multicolumn{8}{c}{Small (S)} \\
    \hline
          & Action Bits & Depth & Num Tree & Max Leaf & lr    & Batch Size & Epoch \\
    \hline
    \grayrow SVM   & 8     &       &       &       &       &       &  \\
    DT$_{EB}$   &       & 4     &       & 1000  &       &       &  \\
   \grayrow  DT$_{DM}$   &       & 4     &       & 1000  &       &       &  \\
    RF$_{EB}$   &       & 4     & 6     & 1000  &       &       &  \\
   \grayrow  RF$_{DM}$   &       & 4     & 6     & 1000  &       &       &  \\
    XGB   &       & 4     & 6     & 1000  &       &       &  \\
   \grayrow   IF   &       &      & 3    &    &      \multicolumn{3}{c}{128 (Num Instance)}  \\
    NB    & 8     &       &       &       &       &       &  \\
   \grayrow KM$_{LB}$   & 8     &       &       &       &       &       &  \\
    KM$_{EB}$   &       & 2     &       &       &       &       &  \\
   \grayrow  KNN   &       & 2     &       &       &   \multicolumn{3}{c}{5 (Num Neighbors)}  \\
    NN    &   binary   &   1(16)     &       &       & 0.01  & 100   & 50 \\
  \grayrow   PCA   & 8     &       &       &       &       &       &  \\
    AE    & 8     &       &       &       & 0.01  & 100   & 50 \\
    \hline
    \multicolumn{8}{c}{Medium (M)} \\
    
    \hline
          & Action Bits & Depth & Num Tree & Max Leaf & lr    & Batch Size & Epoch \\
    \hline
   \grayrow  SVM   & 16    &       &       &       &       &       &  \\
    DT$_{EB}$   &       & 5     &       & 1000  &       &       &  \\
   \grayrow  DT$_{DM}$   &       & 5     &       & 1000  &       &       &  \\
    RF$_{EB}$   &       & 5     & 9     & 1000  &       &       &  \\
   \grayrow  RF$_{DM}$   &       & 5     & 9     & 1000  &       &       &  \\
    XGB   &       & 5     & 9     & 1000  &       &       &  \\
    \grayrow   IF   &       &      & 9     &    &      \multicolumn{3}{c}{128 (Num Instance)}  \\
   NB    & 16    &       &       &       &       &       &  \\
  \grayrow  KM$_{LB}$   & 16    &       &       &       &       &       &  \\
    KM$_{EB}$   &       & 3     &       &       &       &       &  \\
   \grayrow  KNN   &       & 3     &       &       &   \multicolumn{3}{c}{5 (Num Neighbors)}  \\
    NN    &    binary   &  1(32)     &       &       & 0.01  & 100   & 50 \\
  \grayrow   PCA   & 16    &       &       &       &       &       &  \\
    AE    & 16    &       &       &       & 0.01  & 100   & 50 \\
    
    \hline
    \multicolumn{8}{c}{Large (L)} \\
    \hline
          & Action Bits & Depth & Num Tree & Max Leaf & lr    & Batch Size & Epoch \\
    \hline
  \grayrow   SVM   & 32    &       &       &       &       &       &  \\
    DT$_{EB}$   &       & 6     &       & 1000  &       &       &  \\
  \grayrow   DT$_{DM}$   &       & 6     &       & 1000  &       &       &  \\
    RF$_{EB}$   &       & 6     & 12    & 1000  &       &       &  \\
  \grayrow   RF$_{DM}$   &       & 6     & 12    & 1000  &       &       &  \\
    XGB   &       & 6     & 12    & 1000  &       &       &  \\
    \grayrow   IF   &       &      & 12     &    &      \multicolumn{3}{c}{128 (Num Instance)}  \\
   NB    & 32    &       &       &       &       &       &  \\
   \grayrow  KM$_{LB}$   & 32    &       &       &       &       &       &  \\
   KM$_{EB}$   &       & 4     &       &       &       &       &  \\
   \grayrow  KNN   &       & 4     &       &       &   \multicolumn{3}{c}{5 (Num Neighbors)}  \\
    NN    &     binary  &      1(48) &       &       & 0.01  & 100   & 50 \\
  \grayrow   PCA   & 32   &       &       &       &       &       &  \\
    AE    & 32    &       &       &       & 0.01  & 100   & 50 \\
    \hline
    \multicolumn{8}{c}{Huge (H)} \\
    \hline
          & Action Bits & Depth & Num Tree & Max Leaf & lr    & Batch Size & Epoch \\
    \hline
 \grayrow    SVM   & F     &       &       &       &       &       &  \\
    DT$_{EB}$   &       & 30    &       & 100000 &       &       &  \\
 \grayrow    DT$_{DM}$   &       & 30    &       & 100000 &       &       &  \\
    RF$_{EB}$   &       & 30    & 200   & 100000 &       &       &  \\
  \grayrow   RF$_{DM}$   &       & 30    & 200   & 100000 &       &       &  \\
    XGB   &       & 30    & 200   & 100000 &       &       &  \\
    \grayrow   IF   &       &      & 200     &    &      \multicolumn{3}{c}{1280 (Num Instance)}  \\
  NB    & F     &       &       &       &       &       &  \\
  \grayrow  KM$_{LB}$   & F     &       &       &       &       &       &  \\
   KM$_{EB}$   &       & F     &       &       &       &       &  \\
   \grayrow  KNN   &       & F     &       &       &   \multicolumn{3}{c}{5 (Num Neighbors)}  \\
    NN    &  F &    1(48)   &       &       & 0.01  & 100   & 50 \\
 \grayrow    PCA   & F     &       &       &       &       &       &  \\
    AE    & F     &       &       &       & 0.01  & 100   & 50 \\
    \hline\hline
    \end{tabular}%
  \end{adjustbox}
  \vspace{0.5em}
  \caption{Detailed parameters setting for (S)mall, (M)edium, (L)arge model on data plane device/server and (H)uge model size on server. F - Full precision}
  
  \label{tab:parameters_settings}
\end{table}%

\begin{table*}[htbp]
\begin{adjustbox}{width=1\linewidth,center}
  \centering
    \begin{tabular}{ccccccccccccccccccccccc}
    \hline
          & \multicolumn{4}{c}{Iris}      & \multicolumn{6}{c}{KDD99}                     & \multicolumn{6}{c}{AWID3}                     & \multicolumn{6}{c}{Requet} \\
    \hline
          & \multicolumn{2}{c}{Switch (M)} & \multicolumn{2}{c}{Sklearn (M)} & \multicolumn{2}{c}{Switch (M)} & \multicolumn{2}{c}{Sklearn (M)} & \multicolumn{2}{c}{Server (H)} & \multicolumn{2}{c}{Switch (M)} & \multicolumn{2}{c}{Sklearn (M)} & \multicolumn{2}{c}{Server (H)} & \multicolumn{2}{c}{Switch (M)} & \multicolumn{2}{c}{Sklearn (M)} & \multicolumn{2}{c}{Server (H)} \\
    \hline
    Model & ACC   & F1    & ACC   & F1    & ACC   & F1    & ACC   & F1    & ACC   & F1    & ACC   & F1    & ACC   & F1    & ACC   & F1    & ACC   & F1    & ACC   & F1    & ACC   & F1 \\
    \hline
    \grayrow SVM   &  97.78     &    97.81   &  97.78     &    97.81    &       97.17    &  95.15     &  97.17    &  95.15    &   97.17    &  95.15     &  97.17   & 49.28     &  97.00     &  69.23     &    97.00     &  69.23    &   98.13   &  92.03    &    98.13   &   92.03    &    98.13  &  92.03\\
             DT$_{EB}$ &     95.56   &    95.56     &  95.56   &    95.56    &         98.92   &    98.31   &    98.92   &    98.31  &      99.04  &    98.49   &    99.73    &   97.47    &   99.73    &   97.47    &    99.87   &  98.77     &   98.13    &   92.03    &   98.13    &   92.03    &    98.13   & 92.03 \\
    \grayrow DT$_{DM}$ &     95.56   &    95.56     &   95.56   &    95.56    &      98.92   &    98.31   &    98.92   &    98.31    &      99.04  &    98.49   &      98.92   &    98.31     &   99.73    &   97.47    &   99.87   &  98.77  &  98.13 & 92.03   &  98.13 & 92.03  &  98.13 & 92.03  \\
             RF$_{EB}$   &    95.56   &    95.56  &  95.56   &    95.56     &     98.84   &      98.19    &   98.93 &    98.33   &     99.04  &    98.49    &      99.27    &   92.31  &   99.27    &   92.31   &    99.87   &  98.77     &  98.13 & 92.03   &  98.13 & 92.03  &  98.13 & 92.03  \\
    \grayrow RF$_{DM}$   &    95.56   &    95.56    &  95.56   &    95.56   &      98.93     &    98.33   &  98.93     &    98.33     &   99.04  &    98.49      &     99.27    &   92.31     &   99.27    &   92.31    &    99.87   &  98.77  & 98.13 & 92.03   &  98.13 & 92.03  &  98.13 & 92.03\\
             XGB &     97.78     &    97.81     &  97.78     &    97.81   &    98.82   &    98.16      &     98.82   &    98.16   &    99.04   &  98.50  &       99.79  &  98.00     &     99.79  &  98.00     &     99.87   &  98.77     &   98.13    &   92.03    &   98.13    &   92.03 &98.13    &   92.03 \\
    \grayrow IF &    15.56      &    18.18    &   11.11   &    8.48  &  80.31  &  44.54  &     17.77 &    17.51   & 14.55   &  13.67  &    14.40    &     13.37  &  62.39   &  44.23  &   78.08    &  48.00    & NF &   NF   &    23.95  & 14.63  &  25.12 &  16.09  \\
     NB &    95.56   &    95.56   &   95.56    &   95.56    &     96.26    &   93.93  &    95.01    &  91.60   &   95.01    &  91.60    &    85.45   & 59.35      &    85.26   &      59.16 &  85.26   &      59.16    &    91.97   &    78.76   &     91.97   &    78.76   &    91.97   &    78.76 \\
    \grayrow KM$_{LB}$ &    88.89   &    88.19   &  88.89   &    88.19    &   19.51   &   16.34     &     19.51   &   16.34    &    19.51   &   16.34   &       70.97  &    47.5    &      70.97  &    47.57     &      70.97  &    47.57    &   87.06    &   46.54   &   87.06   &    46.54   &    87.06   & 46.54\\
     KM$_{EB}$   &77.78       &   75.93    &   88.89   &    88.19   &   19.55    &    16.37   &    19.51   &   16.34    &     19.51   &   16.34    &       70.97  &    47.5    &     70.97  &    47.57   &       70.97  &    47.57  &     87.06    &   46.54   &   87.06   &    46.54   &    87.06   & 46.54 \\
     
    \grayrow KNN &      80.0    &    66.43    &     100.0  & 100.0     &     40.99  &  40.98      &  99.01    & 98.45      &  99.01    & 98.45 &   97.22     &    49.37   &  99.87   &  98.78    &   99.87   &  98.78     &  90.33    &  64.96    &   99.88 & 99.76  &  99.88 & 99.76   \\
             NN &    93.33   &    93.42   &   95.56    &     95.56    &   75.85    &  71.69     &    99.02   &    98.47   &   99.03    &    98.47   &   96.69    &  62.73    &        99.84   &    98.50   &   99.85    &    98.54  &   93.19    &   48.24    &     98.13   &    92.03 &  98.13   &    92.03 \\
    \hline
          & P1    & P2    & P1    & P2    & P1    & P2    & P1    & P2    & P1    & P2    & P1    & P2    & P1    & P2    & P1    & P2    & P1    & P2    & P1    & P2    & P1    & P2 \\
    \hline
    \grayrow PCA &    100   &   100    &   100    &    100   &  100     &  99.92     &  100     &   100    &     100  &     100  &  100  &     100    &  100  &     100   &  100  &     100   &   100    &   100    &   100    &   100    &    100   & 100 \\
             Auto &   99.94    &   99.95    &   100    &    100   &   99.99    &    100   &     100  &  100     &   100    &    100   &   100    &      100 &  100  &     100    &    100  &     100    &    99.97   &  100     &   100     &     100   & 100 &100  \\
   \hline
    \end{tabular}%
  \end{adjustbox}
  \vspace{0.5em}
  \caption{Evaluation results for (M)edium, (H)uge model functionality in terms of accuracy on different datasets. }
  \label{tab:apdx-functionality}%
  \vspace{-2em}
\end{table*}%

\begin{table}[htbp]
  \begin{adjustbox}{width=1\linewidth,center}
 
\centering

  \begin{threeparttable}
\begin{tabular}{ccccccccccc} 
\hline
\multicolumn{1}{l}{}      & \multicolumn{4}{c}{NASDAQ}  & \multicolumn{6}{c}{Jane Street Market}     \\ 
\hline
\multicolumn{1}{l}{}      & \multicolumn{2}{c}{Switch (M)} & \multicolumn{2}{c}{Sklearn (M)}   & \multicolumn{2}{c}{Switch (M)} & \multicolumn{2}{c}{Sklearn (M)} & \multicolumn{2}{c}{Server (H)}  \\ 
\hline
\multicolumn{1}{l}{Model}    & ACC   & F1    & ACC   & F1         & ACC   & F1          & ACC   & F1        & ACC   & F1    \\ 
\hline
\grayrow SVM & 33.10 & 16.58 & 66.30 & 65.33 & 72.22  &   64.42  & 72.22  &   64.44        & 72.22  &   64.44     \\
DT$_{EB}$ & 89.48 & 89.42 & 89.48 & 89.42 &   72.58  &    67.18    & 72.58  &    67.18  & 75.09 & 71.88          \\
\grayrow DT$_{DM}$ & 89.48 & 89.42 & 89.48 & 89.42 &   72.58   &    67.18    &  72.58   &    67.18 &  75.09 & 71.88  \\
RF$_{EB}$ & 86.21 & 86.14 & 89.68 & 89.64 &  72.62 &  65.91   &  72.62 &  66.43        &  79.94  & 76.63          \\
\grayrow RF$_{DM}$ & 85.14 & 85.06 & 90.02 & 89.94 &     72.62 &  66.43    & 72.62 &  66.43    &  79.94. & 76.63     \\
XGB\tnote{\dag} & 90.25 & 90.20 & 90.26 & 90.21 & 72.50  & 66.80 & 72.50      & 67.24     & 78.73 &  75.61 \\
\grayrow IF\tnote{\dag} & 20.69 & 13.51 & 20.50 & 15.40 & 61.57   &     53.65     &  66.11 &    56.16   &  65.31 &  54.36         \\
NB & 70.94 & 70.46 & 70.90 & 70.41 &   71.70   &    67.17      & 71.64  & 67.26  & 71.64  & 67.26    \\
\grayrow KM$_{LB}$ & 47.31 & 47.66 & 47.31 & 47.66 &  70.42  & 67.87  &  70.42 &  67.87         & 70.42 &  67.87        \\
KM$_{EB}$ & 47.20 & 36.00 & 47.31 & 47.66 & 71.64   &      60.90    & 70.42 &  67.87        &  70.42 &  67.87           \\
\grayrow KNN & 22.94 & 12.96 & 92.41 & 92.40 &  67.21  &    40.22      &  73.68 &   69.63        &  73.68 &   69.63         \\
NN & 49.50 & 47.77 & 92.51 & 92.50 &  64.15   &  58.32  & 72.66  &  67.01         & 72.58 & 67.18  \\
\hline
          & P1    & P2    & P1    & P2    & P1    & P2    & P1    & P2    & P1    & P2    \\
    \hline
    \grayrow PCA & 100 & 100 &    100   &   100    &  100     &   100    &    100   &     100  &  100     &  100      \\
     Auto & 100 & 100 &  100     &    100   &   99.99    &    99.99   &      100     &  100      &    100     &  100        \\
   \hline
\end{tabular}
  \begin{tablenotes}
    \item[\dag] XGB and IF use (S)mall size model in Switch and Sklearn. 
  \end{tablenotes}
  \end{threeparttable}
  \end{adjustbox}
  \caption{Evaluation results for (M)edium model functionality in terms of accuracy on NASDAQ dataset (Stock: AMD) and on Jane Street Market Prediction dataset.
  }
  \label{tab:finance}%
  \vspace{-3em}
\end{table}%

\subsection{Hyperparameter Settings}
Section \ref{sec:eval-modelfunc} provides the functionality tests on the ML models in Planter. For the two datasets CICIDS 2017 (CICIDS) and UNSW-NB15 (UNSW), we use 5-tuple (source and destination port, source and destination IP, and protocol) information as input features. 

The detailed hyperparameter setup for each model is summarized in Table~\ref{tab:parameters_settings}. Hyperparameters related to the converted model size are defined in a gradient scale to differentiate the S/M/L model size. For the huge models, large hyperparameter values are used for full precision accuracy. Other hyperparameters for each model remain as the default values as defined in the \textit{scikit-learn} package. 

\subsection{Evaluation on Additional Datasets} \label{apdix: eval-add-dataset}
As for the datasets, besides the two attack detection datasets presented in the main contents, we also evaluate the accuracy performance on public datasets collected from different application scenarios as presented in Table~\ref{tab:apdx-functionality} and Table~\ref{tab:finance}. Among these six datasets, KDD99 \cite{kddfeatures} and AWID3 \cite{chatzoglou2021empirical} are Intrusion Detection datasets; Requet \cite{requetdataset} is a QoE dataset; Iris \cite{fisher1936useIrisDataset} dataset is a pattern recognition dataset; NASDAQ \cite{nasdaqnasdaq} and Jane Street Market Prediction dataset \cite{janekaggle}  are stock market datasets.

\textbf{Datasets and features} In AWID3 dataset, we input the 5-tuple (source and destination port, source and destination IP, and protocol) information. While in KDD99 dataset, it depicts the traffic flows from another scope and does not provide 5-tuple features, so we choose five basic characteristics for the L4 connection: duration, protocol\_type, service, flag, and land, as input features for each model. Considering the different types of data and use cases might affect the model performance, Requet, Iris, Jane Street Market Prediction and NASDAQ datasets are also tested. Requet is a Quality of Experience (QoE) metric detection dataset collected from video streaming services. The QoE metric like buffer warning is a binary prediction for resource provisioning to avoid streaming stall events. In our case, five features reflecting the streaming sources and states are used: Buffer Progress, Playback Progress, source IP, Playback Quality, and Buffer Valid. Iris is a classic pattern recognition dataset including four features to classify Iris flower type. In our experiments, we use all four features as input to the ML models. As to NASDAQ dataset, we test with add-order messages for one stock. We choose three features (order side, size, and price) in raw messages and use data oversampling to deal with class imbalance. In the Jane Street Market Prediction dataset, five features (42, 43, 120, 124 and 126) among 130 anonymized real stock market data are used to predict buy or sell for each trading opportunity \cite{janekaggle}.

\textbf{Results} The results show that the ML models deployed on the switch with Planter are able to reach the similar performance as on the server. It aligns with the results in section \ref{sec:eval-modelfunc} Table~\ref{tab:functionality}. Most of the models are not that sensitive to the differences among the input data in datasets. Nonetheless, KM$_{EB}$ presents the accuracy loss on the switch than the server in Iris dataset, while NN presents different levels of loss in all the datasets.  


\section{Additional Evaluation Details}

\begin{figure}[htbp]
	\centering
	\includegraphics[width=1\columnwidth]{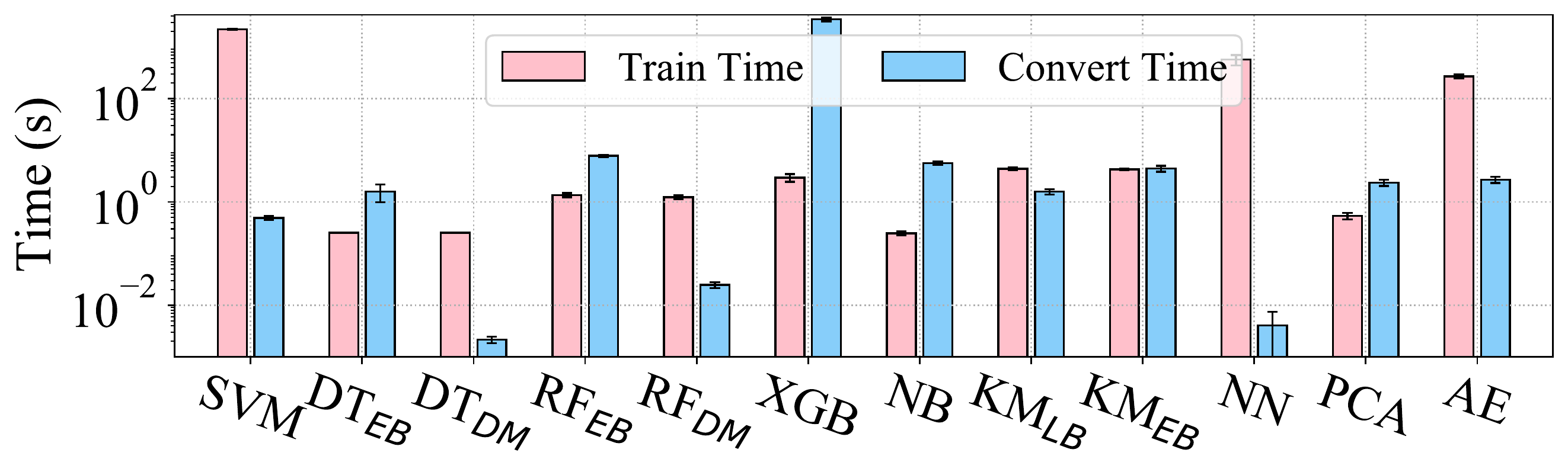}
    \vspace{-2em}
	\caption{Train and convert time of each ML algorithm on top of the attack detection use case.}
    \label{fig:Model_name-Times-M}
	\vspace{-1em}
\end{figure}

The train and convert time (average value of 5 independent trial results) of each ML algorithm in Figure~\ref{fig:Model_name-Times-S} are evaluated with the small models. In addition to that, we also record the time with the medium model size. The small and medium models are defined with the hyperparameters listed in Table~\ref{tab:parameters_settings}. The overall time consumption of each model is similar between the small and medium size. With larger model sizes, the main differences are XGB and KM$_{EB}$ yield longer convert time. It shows that the model size (hyperparameters) affects the number of converted table entries and stages, thereby affecting the convert time of those models sensitive to the model/convert hyperparameters. It corresponds to the results in Figure~\ref{fig:scal1} where XGB and KM$_{EB}$ are sensitive to the hyperparameter change.

 \begin{figure}[htbp]
	\centering
	\begin{minipage}{0.49\linewidth}
	\includegraphics[width=1\linewidth]{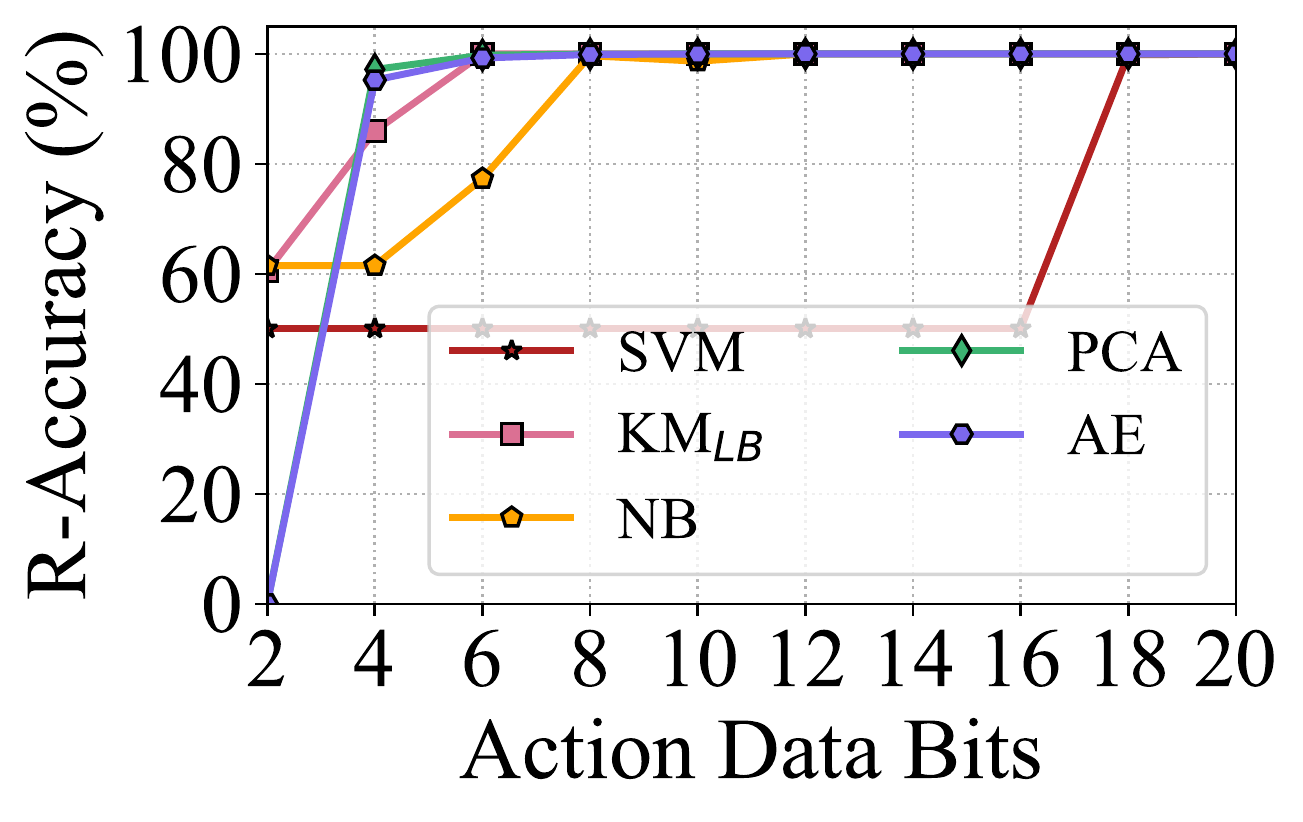}\vspace{-0.5em}\\\centering(a) R-ACC vs. Action Bits
\end{minipage}
	\begin{minipage}{0.49\linewidth}
	\includegraphics[width=1\linewidth]{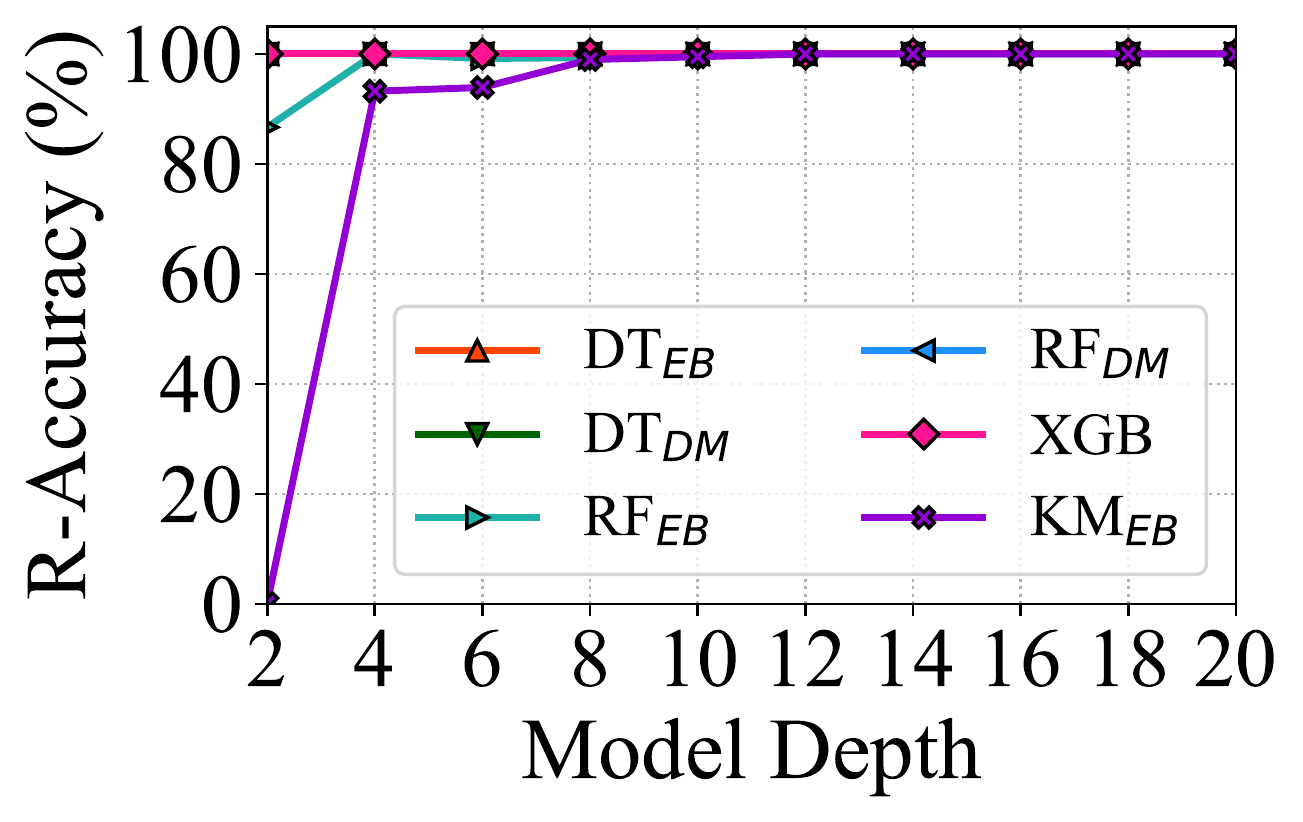}\vspace{-0.5em}\\\centering(b) R-ACC vs. Model Depths.
\end{minipage}
	\vspace{-1em}
	\caption{Relative accuracy between switch and scikit-learn results with two typical model hyperparameters based on CICIDS dataset.}
	\label{fig:acc_cicids}
	\vspace{-0.5em}
\end{figure}

Following the accuracy evaluation in Table~ \ref{tab:functionality}, we study the accuracy performance with different hyperparameter settings in both UNSW and CICIDS datasets. Figure~\ref{fig:relative acc} in the main contents plots the results with UNSW dataset, and Figure~\ref{fig:acc_cicids} portrays the results with CICIDS dataset. Both two figures indicate that the model accuracy on switch can reach the same level as on server as long as the key hyperparameters are set to reach a relatively large converted model size. The difference is that the model might not be able to give accurate results when the model size is relatively small in CICIDS dataset. For instance, KM$_{LB}$, KM$_{EB}$, SVM and NB can only give relatively high accuracy when the converted model is set to a large size.

\section{Ethics and Trust}

The development of AI and \ml{} raises questions of ethics and trustworthiness of the developed system. In this section we extend on these aspects in the context of this work. We distinguish between the development of Planter and the use of Planter.

\paragraph{Development:} The development of Planter complied with all applicable ethical standards of the authors' home institution. No human participants and no personal data were involved in this research. All the datasets used in this research were publicly available and not collected by the authors. 

\paragraph{Use:} The authors acknowledge that use of AI and \ml can be for unethical purposes, and that systems such as Planter can be leveraged by malicious actors, however Planter does not provide such actors new capabilities that are not already available through the use of GPUs and bespoke accelerator cards. Planter focuses on the classification of data, rather than on its training, and does not provide any innovation in training. Same as using \textit{scikit-learn} directly, misconfiguration of Planter can lead to sub-optimal training results. The authors have conducted a thorough evaluation to test the accuracy of classification results, compared with other methods, and inaccuracies in some cases are reported. 

\paragraph{Trust:} The authors follow guidelines proposed in previous works (e.g.~\cite{avin2021filling}) to increase the trustworthiness of Planter. In particular, Planter is made available under an open-source license, and means for the reproducibility of the results presented in this paper are provided.

\end{document}